\documentclass[twocolumn,tighten,astrosymb,trackchanges]{aastex631}
\newcommand{\LJMU}{\affiliation{Astrophysics Research Institute, Liverpool John Moores University, 146 Brownlow Hill, Liverpool L3 5RF, UK}}

\newcommand{\HUBerlin}{\affiliation{Institut f\"ur Physik, Humboldt-Universit\"at zu Berlin, Newtonstr. 15, 12489 Berlin, Germany}}

\newcommand{\Lyon}{\affiliation{Universit\'e Lyon 1, CNRS, IP2I Lyon, UMR 5822, Villeurbanne, France}}
\newcommand{\OKC}{\affiliation{Department of Physics, Oskar Klein Centre, Stockholm University, SE-106 91, Stockholm, Sweden}}
\newcommand{\OKCAstro}{\affiliation{Department of Astronomy, Oskar Klein Center, Stockholm University, SE-106 91 Stockholm, Sweden}}

\newcommand{\CaltechOO}{\affiliation{Caltech Optical Observatories, California Institute of Technology, Pasadena, CA 91125, USA}}

\newcommand{\Caltech}{\affiliation{Cahill Center for Astronomy and Astrophysics, California Institute of Technology, Mail Code 249-17, Pasadena, CA 91125, USA}}

\newcommand{\UCB}{\affiliation{Department of Astronomy, University of California, Berkeley, CA 94720-3411, USA}}

\newcommand{\IoAKavli}{\affiliation{
Institute of Astronomy and Kavli Institute for Cosmology, University of Cambridge, Madingley Road, Cambridge, CB3 0HA, UK}}

\newcommand{\Birmingham}{\affiliation{School of Physics \& Astronomy and Institute for Gravitational Wave Astronomy, University of Birmingham, Birmingham, B15 2TT, UK}}

\newcommand{\Weizmann}{\affiliation{Department of Particle Physics and Astrophysics, Weizmann Institute of Science, Rehovot, Israel}}

\newcommand{\LBL}{\affiliation{Lawrence Berkeley National Laboratory, 1 Cyclotron Road, MS 50B-4206, Berkeley, CA 94720, USA}}

\newcommand{\NCCH}{\affiliation{University of North Carolina at Chapel Hill, 120 E. Cameron Ave., Chapel Hill, NC 27514, USA}}

\newcommand{\NGO}{\affiliation{University of Nova Gorica,
Center for Astrophysics and Cosmology, University of Nova Gorica, Vipavska 11c, 5270 Ajdovščina, Slovenia}}

\newcommand{\BMHH}{\affiliation{School of Physics and Astronomy, University of Birmingham, Birmingham, B15 2TT, UK}}

\newcommand{\Portsmouth}{\affiliation{Institute of Cosmology and Gravitation, University of Portsmouth, Burnaby Rd, Portsmouth PO13FX,UK}}

\usepackage{CJK}
\usepackage{multirow}
\let\tablenum\relax
\usepackage{siunitx}
\usepackage{appendix}
\usepackage{amsmath,amsthm,amsfonts,amssymb,amscd}
\usepackage{booktabs}
\usepackage{tablefootnote}
\usepackage{apjfonts}
\usepackage{lipsum}
\usepackage{changepage}
\usepackage{natbib}
\usepackage{chngcntr}

\let\ts=\thinspace
\newcommand{\one}{\ts {\sc i}}
\newcommand{\two}{\ts {\sc ii}}
\newcommand{\three}{\ts {\sc iii}}
\newcommand{\four}{\ts {\sc iv}}
\newcommand{\five}{\ts {\sc v}}
\newcommand{\six}{\ts {\sc vi}}
\newcommand{\seven}{\ts {\sc vii}}
\renewcommand\ion[2]{#1\,\,{\sc{\romannumeral #2}}}

\definecolor{maroon}{rgb}{0.760,0.118,0.337}

\usepackage{xparse}  

\NewDocumentCommand{\companioncite}{o o m}{%
  {\hypersetup{citecolor=red}%
   \IfNoValueTF{#1}
     {\citet{#3}}
     {\IfNoValueTF{#2}
       {\citet[#1]{#3}}
       {\citet[#1][#2]{#3}}}}%
}

\NewDocumentCommand{\companioncitep}{o o m}{%
  {\hypersetup{citecolor=red}%
   \IfNoValueTF{#1}
     {\citep{#3}}
     {\IfNoValueTF{#2}
       {\citep[#1]{#3}}
       {\citep[#1][#2]{#3}}}}%
}

\def\cm{\mbox{\,cm}}
\def\cm3{\mbox{\,cm$^{-3}$}}

\begin{document}

\title{Follow-up of SN~2025wny I: Space-based Observations of the First Multiply-imaged Superluminous Supernova}


\author[0000-0002-4163-4996]{Ariel~Goobar}
\OKC

\author[0000-0001-5975-290X]{Joel~Johansson}
\OKC

\author[0000-0002-8380-6143]{Edvard~Mörtsell}
\OKC

\author[0000-0003-2456-9317]{Cameron~Lemon}
\OKC
\author[0000-0001-6797-1889]{Steve Schulze} 
\Weizmann

\author[0000-0002-2376-6979]{Suhail~Dhawan}
\Birmingham

\author[0000-0001-6343-3362]{Alice~Townsend}
\Birmingham

\author[0009-0001-6911-9144]{Maggie L.~Li}
\Caltech

\author[0000-0003-3658-6026]{Yu-Jing~Qin}
\Caltech

\author[0000-0003-1710-9339]{Lin~Yan}
\Caltech

\author[0000-0002-4223-103X]{Christoffer~Fremling}
\Caltech
\CaltechOO

\author[0000-0002-5619-4938]{Mansi M. Kasliwal}
\Caltech
\CaltechOO

\author[0000-0002-3389-0586]{Peter~Nugent}
\UCB 
\LBL 

\author[0000-0003-4494-8277]{Graham~P.~Smith}
\BMHH

\author[0000-0003-1546-6615]{Jesper~Sollerman}
\OKCAstro

\author[0000-0002-8977-1498]{Igor Andreoni}
\NCCH

\author[0000-0001-5409-6480]{Nikki~Arendse}
\NGO

\author[0000-0001-5564-3140]{Thomas~E.~Collett}
\Portsmouth

\author[0000-0001-8342-6274]{Jakob~Nordin}
\HUBerlin

\author[0000-0000-0000-0001]{Jacob~Osman~Hjortlund}
\OKC

\author[0000-0001-8472-1996]{Daniel~A.~Perley}
\LJMU
\author[0000-0002-8121-2560]{Mickael~Rigault}
\Lyon

\author[0009-0005-6323-0457]{Stephen Thorp}
\IoAKavli

\author[0000-0003-0733-2916]{Jacob~L.~Wise}
\LJMU

\correspondingauthor{Ariel Goobar}
\email{ariel@fysik.su.se}
\shorttitle{Follow-up of SN~2025wny I: Space-based Observations}
\shortauthors{Goobar et al.}

\begin{abstract}
We present space-based follow-up observations of the superluminous Type~I supernova (SLSN-I) SN~2025wny at redshift $z_{SN}=2.0151$, gravitationally lensed by two galaxies at redshifts $z_{\rm{G1}} = 0.3755$ and $z_{\rm{G2}} = 0.3766$ into five resolved images.
SN~2025wny is the first strongly lensed SLSN discovered and the first galaxy-scale lensed supernova for which both photometric and spectroscopic time-delay measurements are feasible. As such, it opens a new observational window for precision cosmology and the study of stellar explosions near the epoch of peak cosmic star formation.

Our follow-up observations comprise
two epochs of \textit{Hubble Space Telescope} ($HST$) imaging, together with near-infrared imaging and spectroscopy obtained with the \textit{James Webb Space Telescope} ($JWST$). From these data, we measure precise astrometry and multi-band photometry for the five resolved supernova images, the host galaxy, and the two deflecting galaxies. HST provides accurate relative image positions and rest-frame ultraviolet photometry, while JWST delivers complementary near-infrared imaging and spectroscopy probing the rest-frame optical at high signal-to-noise ratio. Together they yield a detailed characterization of both the lensing configuration and the supernova spectral energy distribution over a broad wavelength range.

The data presented here provide the observational foundation for the accompanying analyses of the supernova properties, lens modeling, and time-delay cosmography, including the astrometric, photometric, and spectroscopic information required to measure $H_0$.
\end{abstract}


\section{Introduction}

\subsection{Strongly Lensed Supernovae as Probes of Astrophysics and Cosmology}
Strongly gravitationally lensed supernovae (glSNe) are emerging as powerful probes of both stellar explosions and precision cosmology. On the astrophysical side, the magnification provided by the foreground lens can boost intrinsically faint or distant events above the thresholds required for detailed photometric and spectroscopic follow-up, offering an efficient route to studying the physics of individual supernovae that would otherwise be too faint for such scrutiny. On the cosmological side, time delays between multiple supernova images provide a direct route to measuring the Hubble constant, $H_0$, largely independent of the traditional local distance ladder and early-Universe probes. As the discrepancy between late- and early-Universe determinations of $H_0$, known as the ``Hubble tension'' \citep[see][and references therein]{2021CQGra..38o3001D}, has persisted and grown in significance, time-delay cosmography has become an increasingly important independent avenue for testing the standard cosmological model \citep[see][for a recent review]{2024SSRv..220...48B}. Compared to lensed quasars, supernovae offer several important advantages for time-delay measurements, including predictable light-curve evolution, reduced microlensing systematics, and the possibility of spectroscopic phase matching between multiple images \citep{suyu_2024a}.
Since the discovery of the core-collapse supernova SN Refsdal \citep{kelly_2015,Kelly_2016}, the number of known glSNe has increased rapidly. These discoveries include cluster-lensed events
\citep{rodney_2021,chen_2022,pierel_2024,2026arXiv260104156C,2026arXiv260411882D2026arXiv260411882D}, as well as galaxy-scale lensed Type~Ia supernovae discovered in wide-field time-domain surveys \citep{goobar_2017,goobar_2023}; see \citet{Goobar_2025} for a recent review. More recently, two strongly lensed supernovae originating from explosions of massive stars have been identified: SN~2025mkn \citep{Lemon+26} and the subject of this study, the superluminous supernova SN~2025wny \citep{2025ApJ...995L..17J,Taubenberger_wny}.

\subsection{SN~2025wny and This Focus Issue}
\begin{figure*}[!ht]
    \centering
    \includegraphics[width=\textwidth]{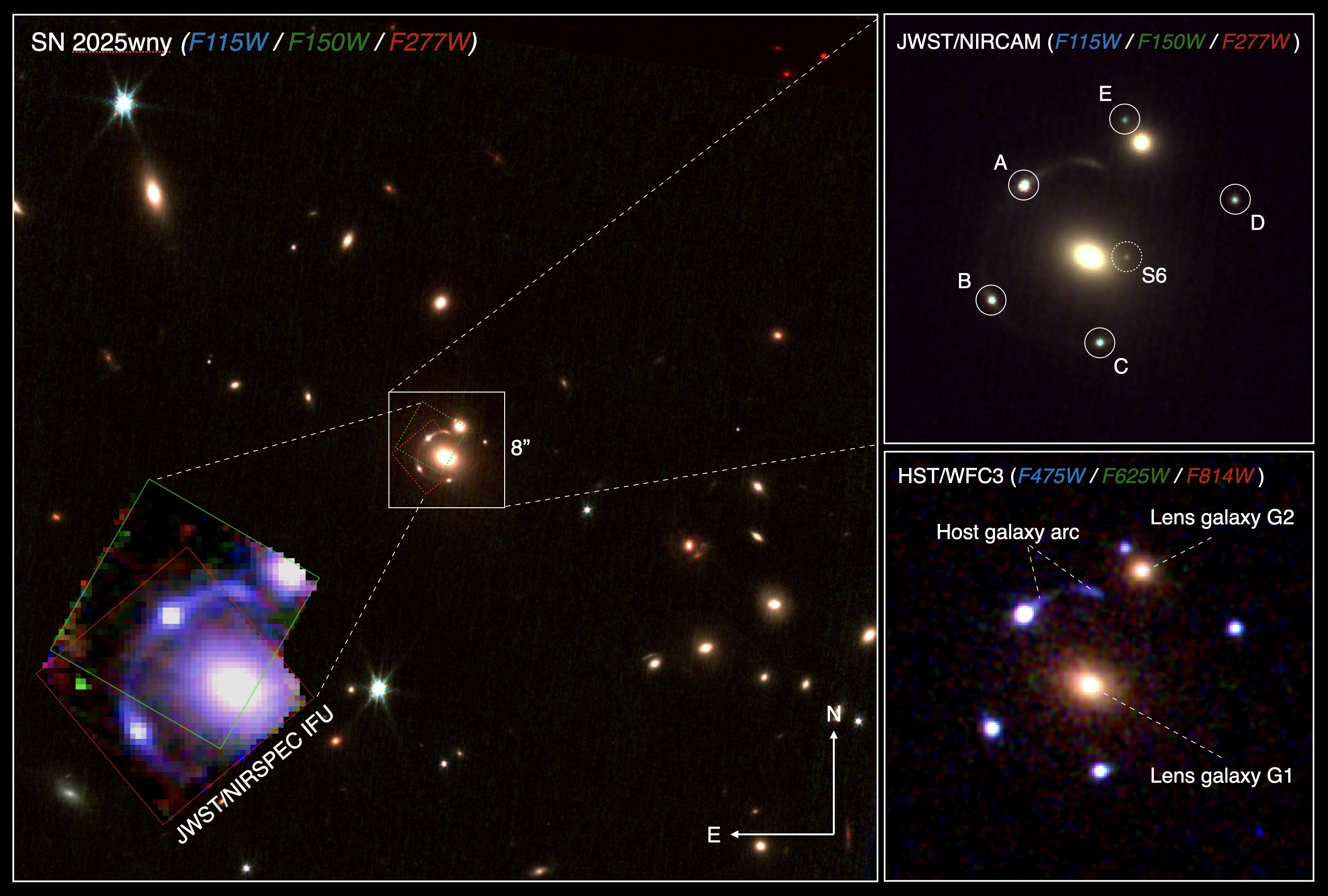}
\caption{Overview of the strongly lensed superluminous supernova SN\,2025wny and its
environment. \textit{Left:} Color composite JWST image of the full $1\arcmin \times
1\arcmin$ field, constructed from NIRCam F115W (blue), F150W (green), and F277W (red).
The inset at the bottom left indicates the two NIRSpec IFU pointings from programs
GO~5564 (red outline) and DDT~12510 (green outline). \textit{Top right:} Zoomed
$8\farcs0\times8\farcs0$ color composite from the same \textit{JWST}/NIRCam filters,
showing the five resolved lensed images of the supernova (A--E) and the additional
unidentified point source S6, each marked with circular apertures.
The two foreground deflecting galaxies, hereafter G1 and G2 \companioncitep[at  $z_{G1} = 0.3755$, and $z_{G2} = 0.3766$, see][]{Johansson2026},
have a projected separation smaller than the Einstein radius of the system, causing
them to act as a single compound lens \companioncitep[see][for further details]{Mortsell2026}. Image~A is the brightest and most magnified; image~D is the most
offset from G1 and observed at a significantly different supernova phase 
\companioncitep[see also][]{Johansson2026}. \textit{Bottom right:} $HST$/WFC3 color composite
constructed from the F475W (blue), F625W (green), and F814W (red) exposures obtained
in Epoch~1 (2025 October 20), clearly resolving the two deflecting lens galaxies
(labeled Lens galaxy~1 and Lens galaxy~2 in this panel, i.e., G1 and G2) and the
brightest arcs of the lensed host galaxy.}    
    \label{fig:pretty_pic}
\end{figure*}

SN~2025wny is the first strongly lensed superluminous supernova discovered, the highest-redshift multiply-imaged supernova identified in a galaxy-scale lens to date, and the first galaxy-scale lensed supernova suitable for precision time-delay cosmography.
Through routine monitoring of the Northern sky by the Zwicky Transient Facility \citep{Bellm+2019, Graham+2019}, ZTF25abnjznp, aka SN~2025wny was reported to the Transient Name Server (TNS\footnote{https://www.wis-tns.org/}) on September 03, 2025. The earliest ZTF $r$-band detection dates back to August 23, 2025. SN~2025wny had also been reported independently to the TNS by the Gravitational-wave Optical Transient Observer (GOTO; \citealt{2022MNRAS.511.2405S}).
Multiple SN images were first resolved from the Liverpool Telescope \citep{2025TNSAN.296....1W} and the transient was spectroscopically classified, along with its redshift, $z_{SN} = 2.01 \pm 0.005$, by \cite{2025TNSAN.306....1J} as a hydrogen-poor superluminous supernova (SLSN-I) based on its blue continuum and lack of hydrogen features.
The supernova lies in close angular proximity to a pair of massive galaxies (which we will refer to as G1 and G2), surrounded by bluer sources that had already been flagged as a likely strong lensing system by \cite{2020A&A...644A.163C}. For one of the galaxies (G1), spectroscopy from the first data release of the Dark Energy Survey Instrument (DESI) was available,
yielding $z_{\rm{G1}} = 0.3754 \pm  0.0001$ \citep{2025arXiv250314745D}. A slight revision of the redshift using Keck observations is provided by 
\companioncite{Johansson2026}.
What singles out SN~2025wny are the wide image separations, large differences in the image arrival times \companioncitep{Johansson2026,Townsend2026} and the rich astrometric data set for lens modeling, including five supernova images and host galaxy arcs, all reported in this work.  The combination of high intrinsic luminosity, large lensing magnification, and well-resolved multiple images makes the SN~2025wny system not only uniquely powerful for precision cosmography, but also for studying the physics of superluminous explosions.
SLSNe constitute the most luminous class of stellar explosions, reaching peak absolute magnitudes of $M \lesssim -21$ mag. SN~2025wny was classified as a hydrogen-poor event (Type~I SLSN) \citep{2025TNSAN.306....1J}, a relatively homogeneous observational class characterized by blue continua near maximum light, broad absorption features associated with intermediate-mass elements, and long characteristic timescales. Their extreme luminosities make them promising probes of massive-star explosions at high redshift, although even SLSNe become observationally challenging beyond $z \gtrsim 2$, where rest-frame ultraviolet emission is shifted into the near-infrared.
In this work, we present the space-based follow-up campaign of SN~2025wny conducted with the \textit{Hubble Space Telescope} ($HST$) and the \textit{James Webb Space Telescope} ($JWST$). The primary goals of these observations were to spatially resolve the supernova images from the foreground lensing galaxies, obtain precise rest-frame ultraviolet and optical photometry, and secure high signal-to-noise near-infrared spectroscopy.
Table \ref{tab:obs_summary} summarizes the space-based follow-up campaign, including JWST and $HST$ observations. Figure \ref{fig:pretty_pic} shows an overview of the observations.
This paper focuses on the observations, data reduction, astrometric measurements, and initial astrophysical characterization of the system. Detailed lens modeling, magnification estimates, time-delay measurements, spectrophotometric analysis, host-galaxy properties, and survey-rate predictions are presented in a series of companion papers. \companioncite{Mortsell2026} model the lens system -- comprising two elliptical power-law mass distributions plus an external shear component -- using the $HST$ and $JWST$ imaging presented here to fit the supernova image positions and the lensed host-galaxy surface brightness, deriving the system's Einstein radii and, after accounting for microlensing by stars in the lens galaxies, a posterior on the total magnification. \companioncite{Johansson2026} present spatially resolved spectroscopy of images A--E spanning several months and measure the relative time delays between images from the temporal evolution of broad absorption-line minima. \companioncite{Townsend2026} derive independent, photometric time delays from multi-band ($griz$) light curves obtained with the Gemini North Telescope, Large Binocular Telescope, Liverpool Telescope, Palomar 60-inch Telescope, and the Fraunhofer Telescope at Wendelstein Observatory, using scene-modelling photometry to deblend the lensed images. Combined with the lens model of \companioncite{Mortsell2026}, the spectroscopic and photometric delays yield independent, complementary determinations of $H_0$. \companioncite{Li2026} present densely sampled rest-frame UV-to-optical photometry and spectroscopy extending to $+80$\,d post-peak from JWST, Keck, the VLT, Gemini, and other facilities, characterizing the supernova's peak luminosity, several unusual spectral features, and the implications for its power source. \companioncite{Qin2026} study the properties of the host galaxy of SN~2025wny, including the stellar mass and properties of the circumstellar and interstellar medium, also in relation to lower redshift SLSNe-I. Finally, \companioncite{Hjortlund2026} present forward simulations of the Zwicky Transient Facility survey to estimate the expected rate of lensed SLSNe-I over seven years of operations, assess whether the discovery of SN~2025wny is consistent with expectations, and quantify how selection effects bias its inferred magnification and intrinsic luminosity relative to the underlying SLSN-I population.


\section{JWST Observations}
We obtained near-infrared imaging and integral-field spectroscopy with JWST \citep{Gardner2023} as part of the
Cycle~3 General Observer program GO~5564 (PI: A.~Goobar), see Table~\ref{tab:obs_summary}. Both
observations were executed as non-disruptive Target-of-Opportunity
(ToO) triggers centered at the target's equatorial
coordinates
$\alpha = 07^{\mathrm{h}}16^{\mathrm{m}}34\fs5053$,
$\delta = +38\arcdeg21\arcmin06\farcs97$ (J2000.0). A summary of the
instrument configurations is given in Table~\ref{tab:jwst_obs}. The observations were carried out 87 and 90 observer-frame days after the first ZTF detection, corresponding to
about a month in the SN restframe. 
Additional spectroscopic observations were carried out just over 200 days after first detection, obtained as part of Cycle~4 Director's Discretionary Time program
DDT~12510 (PI: M.~Li). A composite picture of the obtained  imaging of the system is shown in Figure \ref{fig:pretty_pic}, while the individual filter images, including, galaxy subtractions are shown in Appendix A, Figures \ref{fig:hst_475}--\ref{fig:jwst_277}. The inferred astrometry of the five SN images A--E, the two deflecting galaxies G1 and G2, as well as an additional sixth point source of unknown origin, S6, are summarized in Table \ref{tab:astrometry}, and the photometry of the point sources in Table \ref{tab:photometry}.

\begin{deluxetable*}{lclllc}
\tablecaption{Summary of space-based observations of SN\,2025wny.\label{tab:obs_summary}}
\tablewidth{0pt}
\tablehead{
  \colhead{Date (UT)} &
  \colhead{MJD} &
  \colhead{Program} &
  \colhead{Instrument} &
  \colhead{Filter(s) / Disperser} &
  \colhead{Purpose}
}
\startdata
2025 Oct 20  & 60968 & HST GO\,17611 (Ep.\,1) & WFC3/UVIS & F475W, F625W, F814W, F160W      & Optical and NIR photometry \\
2025 Nov 19  & 60998 & JWST GO\,5564          & NIRSpec IFU & G140M/F100LP                    & IFU spectroscopy (images A, B, G1, S6) \\
2025 Nov 22  & 61001 & JWST GO\,5564          & NIRCam      & F115W, F150W, F277W             & NIR imaging \\
2025 Nov 30  & 61009 & HST GO\,17611 (Ep.\,2) & WFC3/UVIS & F475W, F625W, F814W, F160W       & Optical and NIR photometry \\
2026 Mar 14  & 61113 & JWST DDT\,12510        & NIRSpec IFU & G140M/F100LP,                   & IFU spectroscopy (image A, G2) \\
             &       &                        &             & G235M/F170LP,                   & \\
             &       &                        &             & G395M/F290LP                    & \\
\enddata
\end{deluxetable*}

\subsection{NIRCam Observations}\label{sec:nircam}

Near-infrared imaging was acquired with the Near Infrared Camera
\citep[NIRCam;][]{Rieke2023} on 2025 November 22 at 21:53:26~UT,
 providing high-resolution photometry corresponding to rest-frame optical wavelengths. These data further resolve the supernova images, its host galaxy, and the foreground lens system.

The
observation employed both NIRCam modules, without mosaicking. The dual-channel observations were carried out in two filter pairs
that combine SW and LW channels, through the dichroic:
\textit{F115W}+\textit{F277W} and \textit{F150W}+\textit{F277W}. 
The \textit{F115W} and
\textit{F150W} SW filters, with a platescale of 0\farcs031~pixel$^{-1}$ sample the rest-frame ultraviolet/optical
of the supernova,
whereas the \textit{F277W} LW filter (0\farcs063~pixel$^{-1}$) constrains the host and lens
contributions and provides high-precision astrometry of the
multiply-imaged system.
We adopted the
\texttt{INTRAMODULEBOX} primary dither pattern with two primary
positions, combined with a 2-position \texttt{SMALL-GRID-DITHER}
subpixel pattern, for a total of four dither positions per filter
pair. This enabled us to recover Nyquist-sampled point-spread functions, reject bad pixels, and bridge the inter-module and inter-chip gaps.

All exposures used the \texttt{BRIGHT1} readout pattern with
$N_{\mathrm{groups}}=3$ and $N_{\mathrm{int}}=2$, yielding an
effective on-source time of $\sim$118~s per dither $\times $ four dithers, resulting in $\sim$473~s
per filter pair. The total NIRCam science duration was 946~s.

The choice of \texttt{BRIGHT1} (rather than a deeper readout pattern
such as \texttt{MEDIUM8} or \texttt{DEEP8}) reflects the relatively
bright magnitudes of the resolved supernova images while
still providing the dynamic range needed to detect faint
host-galaxy and lensed-arc features. 

\label{sec:nircam-reduction}

The NIRCam imaging was reduced with the official
JWST Calibration Pipeline \citep{Bushouse2025}, incorporating
targeted modifications motivated by the well-known 1/f
correlated-noise striping that affects raw NIRCam ramps and
propagates into the Stage 1 (\texttt{Detector1Pipeline}) products as
horizontal and vertical banding across each detector. At the surface brightness
levels of the faint lensed arcs and inter-image regions of
SN 2025wny, this striping is comparable to the photon-noise floor
and would otherwise dominate the local background under the
multiple SN images, biasing the point-spread-function-fitting
photometry reported in Table~\ref{tab:photometry}.

We therefore processed the \texttt{\_uncal.fits} files through
Stage~1 of the \texttt{calwebb\_detector1} pipeline in the standard
configuration (including superbias subtraction, reference-pixel
correction, linearity correction, dark subtraction, jump detection,
and ramp fitting) to produce \texttt{\_rate.fits} count-rate images.
Between Stages~1 and~2, we
then applied the \texttt{remstriping.py} routine from the CEERS
NIRCam reduction
toolkit\footnote{\url{https://github.com/ceers/ceers-nircam}}
\citep{Bagley2023,Finkelstein2023}, which performs an iterative,
source-masked sigma-clipped row-and-column median subtraction on
each amplifier of each \texttt{\_rate.fits} frame. This
implementation was  developed to mitigate observations
where $1/f$ residuals limit the achievable surface-brightness
depth. It  masks bright sources prior to the median subtraction so that
real astrophysical flux is not absorbed into the stripe model, and
we found it to remove the banding to the detector noise floor in
our data while leaving the photometry of the resolved SN images
unbiased. 

The de-striped
count-rate images were then passed through Stage~2
and Stage~3, 
using the standard parameter values for all steps except the final output pixel scale, which we set to $0\farcs015$~pixel$^{-1}$ (SW) and $0\farcs030$~pixel$^{-1}$ (LW) to better sample the PSF given our four-point dither pattern. Calibration files from the Calibration Reference
Data System (CRDS) were used.
\subsection{NIRSpec IFU Spectroscopy}\label{sec:nirspec}

Two epochs  of Integral-field spectroscopy were obtained with the Near Infrared
Spectrograph \citep[NIRSpec;][]{Jakobsen2022} Integral Field Unit
\citep[IFU;][]{Boeker2022}.
The IFU has a $3\farcs0\times3\farcs0$ field of
view, sampled by $30\times30$ spaxels of $0\farcs1$ each.
For the first visit, on 2025 November 19,
the FoV encloses two of the resolved multiple images of the lensed system, A and B.

We employed a single disperser--filter combination, the
medium-resolution \textit{G140M} grating with the \textit{F100LP}
order-separation filter, providing a nominal spectral coverage of
$0.97$--$1.89~\micron$ at resolving power $R\simeq1000$. The wavelength range
covers the rest-frame ultraviolet and optical features essential for comparison with SLSNe found at lower redshifts.   

To suppress correlated read noise and $1/f$ detector noise that
limit faint-source spectroscopy, we adopted the improved reference
sampling and subtraction (IRS$^{2}$) readout, specifically the
\texttt{NRSIRS2RAPID} pattern, with $N_{\mathrm{groups}}=85$ and
$N_{\mathrm{int}}=1$ per exposure. Spatial sampling was achieved
with the \texttt{4-POINT-DITHER} pattern, yielding four exposures
at independent IFU pointings for cosmic-ray and bad-pixel rejection.
The total NIRSpec
science duration was 5019~s ($\sim$1.4~hr). 

The DDT program 12510 led by \companioncite{Li2026} was carried out on March 14, 2026 with a slightly different pointing, shown in Figure \ref{fig:pretty_pic}. The IFU was centered at the brightest lensed image of the supernova, Image A, and included the secondary deflecting galaxy G2, but not image B. The pointing also included an extended host-galaxy arc. Three medium-resolution
disperser--filter pairs were used: \textit{G140M}/\textit{F100LP}, i.e., same as for GO 5564, but also \textit{G235M}/\textit{F170LP}
($1.66$--$3.17~\micron$), and \textit{G395M}/\textit{F290LP}
($2.87$--$5.27~\micron$), each providing a nominal resolving power
of $R\simeq1000$. These observations were mainly  used to study the spectral evolution of SN~2025wny as presented in the accompanying paper by \companioncite{Li2026}. Hence, we defer to that paper for further details.

\subsection{$HST$ Observations}\label{sec:hst}

Complementary optical and near-infrared imaging of SN~2025wny was
obtained with the Wide Field Camera~3
\citep[WFC3;][]{Kimble2008,Dressel2023} aboard $HST$ under Cycle~31
General Observer program GO~17611 (PI: A.~Goobar), the
{$HST$} companion to JWST program GO~5564
(\S\ref{sec:nircam}, \S\ref{sec:nirspec}). The program consisted of
two single-orbit visits at distinct epochs: Visit~1 (Epoch~1)
executed on 2025 October~20 at 18:03~UT, and Visit~2 (Epoch~2) on
2025 November~30 at 13:50~UT, approximately 60 and 100 days (observer frame)  after
the first detection. The epoch spacing was chosen to constrain the
supernova flux ratios at two well-separated phases, enabling
cross-checks of the spectroscopic time-delay measurement and
characterization of potential differential dust extinction along the
multiple sight-lines through the lens. The two visits (see Table~\ref{tab:obs_summary}) used identical
instrument configurations, summarized in Tables \ref{tab:obs_summary} and ~\ref{tab:hst-obs}.

In each visit, four filter observations were obtained through the
WFC3 dichroic, three with the UVIS channel and one with the IR
channel. The UVIS exposures employed the $512\times512$ pixel
\texttt{UVIS2-C512C-SUB} subarray ($\sim20\arcsec\times20\arcsec$
field of view at the native UVIS plate scale of
$0\farcs04$~pixel$^{-1}$), through the wide-band filters
\textit{F475W} ($42$~s), \textit{F625W} ($13$~s), and
\textit{F814W} ($20$~s).  A post-flash level of $14~e^{-}$ was
applied to each UVIS exposure to mitigate charge-transfer
inefficiency at low background levels. The IR exposure used the
\texttt{IRSUB512} subarray ($512\times512$ pixels,
$\sim64\arcsec\times64\arcsec$ at the native IR plate scale of
$0\farcs13$~pixel$^{-1}$), through the wide-band \textit{F160W}
filter, in \texttt{MULTIACCUM} mode with the \texttt{SPARS25}
sample sequence and $N_{\rm SAMP}=10$ non-destructive readouts
(total integration time $\sim199$~s). A \texttt{POS-TARG} offset of
$+10\arcsec$ along the detector $Y$-axis was applied to the IR
exposure to position the target away from the known
persistence-prone region of the IR detector.

To Nyquist-sample the WFC3 point-spread function and reject bad
pixels, the exposures were dithered using a 3-point
linear dither pattern: \texttt{WFC3-UVIS-DITHER-LINE-3PT} with a
point spacing of $0\farcs135$ at orientation $46\fdg84$ for the
UVIS exposures, and \texttt{WFC3-IR-DITHER-LINE-3PT} with a point
spacing of $1\farcs305$ for the IR exposure.




The data were retrieved from the Mikulski Archive for Space Telescopes and processed using standard \texttt{AstroDrizzle} procedures \citep{2012AAS...22013515H}. PSF-fitting photometry was employed to minimize contamination from the lensing galaxies.

\section{Observational Properties}
\begin{figure*}[htp]
    \centering
\includegraphics[width=0.9\textwidth]{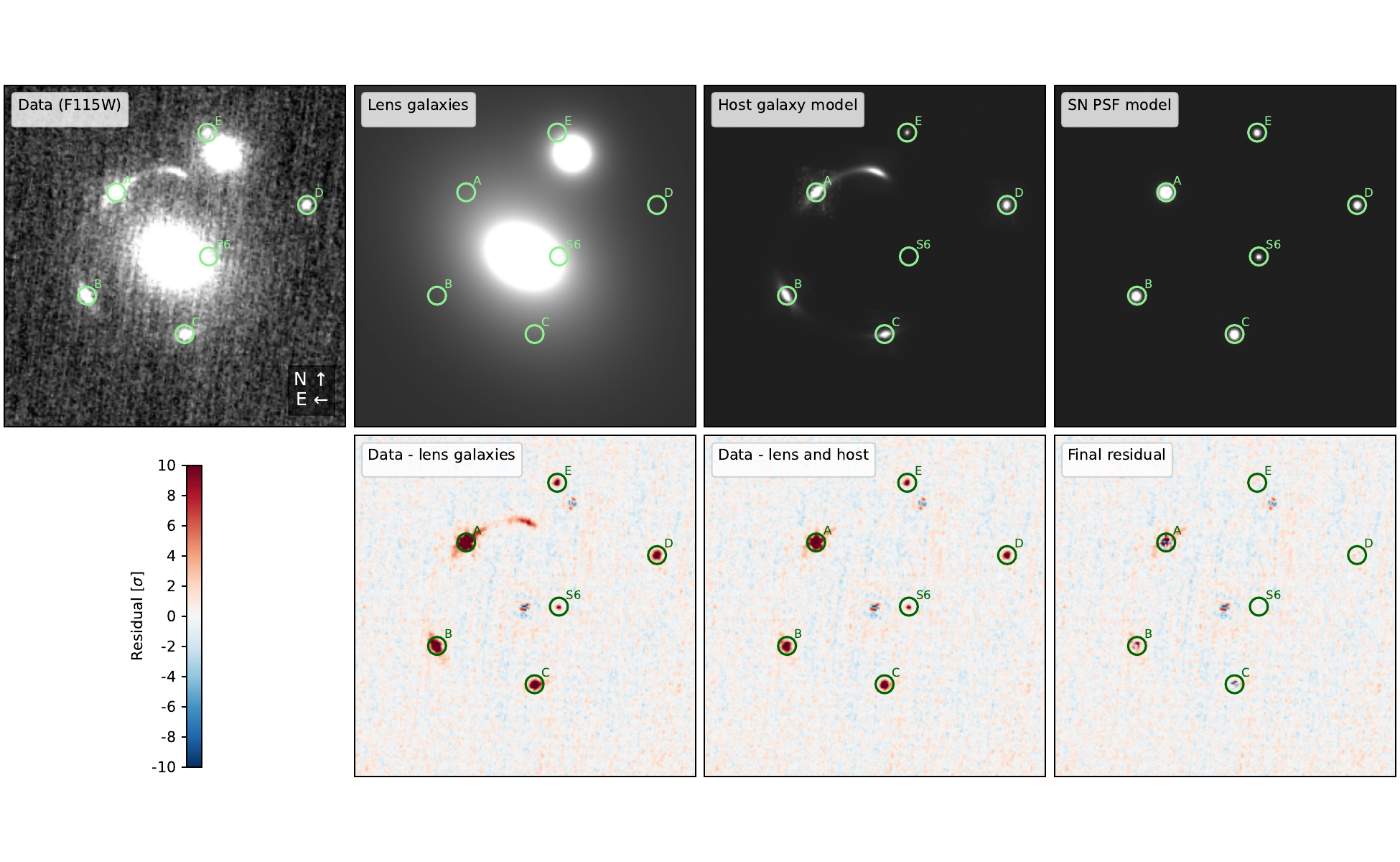}  
     \caption{Image decomposition of SN~2025wny in the \textit{JWST}/NIRCam F115W band. 
Each panel is $10\farcs0 \times 10\farcs0$ with north up and east to the left. 
\textit{Top row, left to right:} The reduced science image; the best-fit 
foreground lens galaxy model (two S\'{e}rsic profiles); the best-fit host 
galaxy model; and the best-fit PSF model for the five lensed SN images 
(A--E) and point source S6. \textit{Bottom row, left to right:} The 
residual after subtracting the lens galaxy model; the residual after 
additionally subtracting the host galaxy model; and the final residual 
after subtracting all components. Residuals are shown in units of the 
local noise $\sigma$, with the color scale saturating at $\pm10\,\sigma$.}
    \label{fig:jwst_115}
\end{figure*}

We performed photometry of the individual lensed supernova images using PSF fitting after modeling and subtracting the foreground lens galaxies and the lensed host-galaxy emission. The same procedure was applied to the $HST$/WFC3 and JWST/NIRCam imaging. 

For each dataset, we modeled the PSF with a Moffat profile fitted to nearby isolated field stars. The foreground lens galaxies were modeled with a compound Sérsic model convolved with the image PSF, together with a constant background component. 
The extended lensed host-galaxy emission was modeled separately \companioncitep{Mortsell2026} and resampled onto the WCS and pixel grid of each science image before subtraction. 
After subtraction of the lens and host galaxies, we simultaneously fitted PSF components to the supernova images A--E and S6. Their positions were initialized from the known sky coordinates, while the individual PSF amplitudes were fitted independently and converted to AB magnitudes. For the astrometric measurements, the source centroids were additionally allowed to vary around their initial positions.

\autoref{fig:jwst_115} shows an example of the photometry steps for JWST/NIRCam F115W data:  the best-fit 
foreground lens galaxy model (two S\'{e}rsic profiles); the best-fit host 
galaxy model; and the best-fit PSF model for the five lensed SN images 
(A--E) and point source S6. The bottom row shows the residuals after subtracting the lens galaxy model; the residual after 
additionally subtracting the host galaxy model; and the final residual  after subtracting all components. (For the $HST$ and remaining JWST filters, see Figures \ref{fig:hst_475} -- \ref{fig:jwst_277}).

Photometric uncertainties were estimated by combining the formal PSF-fit uncertainty and the local scatter from artificial-source injections. Artificial sources were injected around each SN in the subtracted images and recovered using the same PSF-fitting procedure, with the resulting scatter accounting for local background and galaxy-subtraction residuals. The same simulations were used to estimate limiting magnitudes for low-significance measurements.

We additionally measured the lensed host-galaxy flux using aperture photometry at a fixed sky position in the foreground-lens-subtracted images, adopting a $0\farcs3$ aperture radius and a local background annulus spanning $0\farcs4$--$0\farcs5$. Table~\ref{tab:astrometry} summarizes the space-based astrometry, the extracted photometry for the point sources is shown in 
Table~\ref{tab:photometry}, while 
Table~\ref{tab:flux_ratios_by_image} lists the image flux ratios relative to image A. Finally, 
Table~\ref{tab:deflector_photometry} shows the measured photometry for the two deflecting galaxies.

\subsection{Photometric properties}\label{sec:flux_ratios}
\begin{figure*}[htp]
    \centering
    \includegraphics[width=0.8\linewidth]{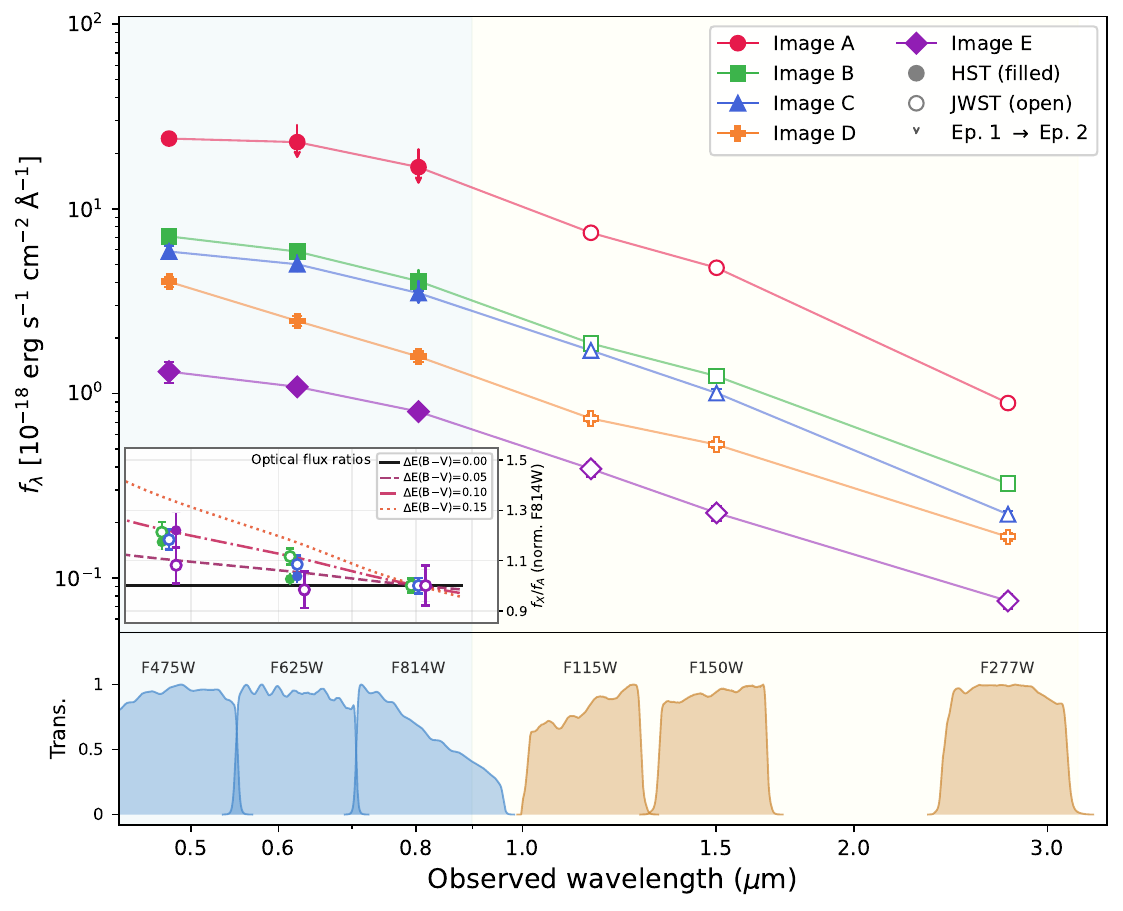}
    \caption{Spectral energy distributions of the five lensed images of SN\,2025wny
    (A--E) constructed from \textit{HST} WFC3-UVIS (F475W, F625W, F814W; filled symbols)
    and \textit{JWST} NIRCam (F115W, F150W, F277W; open symbols) photometry.
    For the two-epoch \textit{HST} bands the plotted flux densities are
    inverse-variance-weighted means over both visits
    (MJD\,60969 and 61010); the \textit{JWST} points correspond to a
    single epoch (MJD\,$\approx$\,61002).
    The light-blue and yellow shading delineate the \textit{HST} optical and NIR/IR wavelength ranges, respectively.
For an achromatic macro lens and identical intrinsic source spectra at the sampled phases, the SEDs would differ only by a multiplicative normalization. Deviations from this would indicate chromatic differences between images, arising from
differential dust extinction in the lens galaxy 
    and/or intrinsic color evolution driven by the image-to-image time delays. The latter is certainly the case for image D which has a 
    considerable phase difference compared to the other SN images, see \companioncite{Johansson2026}.
    Connecting lines are drawn to guide the eye. Image A shows stronger flattening in the optical filters compared to images B, C and E at similar phases, possibly explained by a differential extinction of E(B-V) $\lesssim 0.1$ mag in the deflecting galaxy. 
Arrows at the $HST$ filter positions indicate the change in $f_\lambda$ between the two $HST$ epochs (2025 October~20 and 2025 November~30), separated by 41 days in the observer frame; all five images faded between the two epochs, with image~A showing the largest decline.    
    The lower panel shows the normalized filter transmission curves. 
    {\em Inset}: $HST$ optical flux ratios
  $f_X/f_A$ for images B, C, and E, normalized to their F814W value and plotted against the same (aligned)
  observed-wavelength axis as the main panel. Overplotted are Milky-Way like dust \cite{Fitzpatrick1999} extinction curves
  ($R_V=3.1$) for a dust screen at the lens redshift $z_{\rm lens}=0.375$, anchored at F814W, for differential
  reddening $\Delta E(B-V)=0.00,0.05,0.10,0.15$ mag relative to image A; filled and open symbols correspond to
  the first and second $HST$ epochs. The data are consistent with $\Delta E(B-V)\lesssim 0.1$ mag of differential
  extinction between the images, indicating only modest image-to-image color differences.
    }
    \label{fig:SED_images}
\end{figure*}
Image A is consistently the brightest, and thus most magnified image. 
Figure \ref{fig:SED_images} shows the Spectral Energy Distributions extracted from the $HST$ and JWST photometric observations tabulated in Table ~\ref{tab:photometry}.
The flux ratios between the five lensed images encode the relative
magnifications of the macro-lens model, modulated by image-to-image time delays, differential
extinction, galaxy contamination,  and (potentially) chromatic
microlensing \companioncitep[see][for a discussion of possible microlensing of image A, in particular]{Townsend2026}. We report the ratios as $f_X / f_A$ relative to
image~A  and propagating the photometric uncertainties on each magnitude in
quadrature, shown in
Table~\ref{tab:flux_ratios_by_image}. For the four \textit{HST} bands, observed at two epochs
each, the per-epoch ratios are combined using inverse-variance
weighting. The three $JWST$/NIRCam bands were observed in a single
visit (MJD~$\approx 61001.96$) and contribute a single measurement
each. Using the $HST$ and $JWST$ photometry, we measure colors for each SN image. If interpreted entirely as differential extinction, the optical color differences correspond approximately to $\Delta E(B-V) \approx 0.07$--$0.09$\,mag relative to images B, C and E is broadly consistent with the optical $HST$ data suggesting reddening
along image~A's sightline,  as shown in the inset of Figure~\ref{fig:SED_images}. This reddening corresponds to a dimming of $\Delta m_{g,r}$ of 0.4 and 0.3 mag, respectively. Unlike the case for image D, images A, B, C and E have relatively small phase offsets \companioncitep[see][]{Johansson2026} and can hence be assumed to have comparable SEDs from which the color differences can be inferred. Besides some potential intrinsic color difference between the images due to their different phases, the measurement accuracy is limited by possible unsubtracted galaxy contamination (see Figures \ref{fig:hst_475}--\ref{fig:jwst_277})  and hence should be taken as only preliminary.


Among the seven bands, F160W and F150W sample essentially the same
rest-frame wavelength ($\lambda_{\mathrm{rest}} \approx 5100\,$\AA \,  at
$z = 2.01$) using two different instruments. The \textit{HST}/WFC3
F160W flux ratios are inverse-variance averages over the two
{$HST$} epochs, while the $JWST$/NIRCam F150W ratios come from
the single $JWST$ visit, falling between the two {HST} epochs
(MJD~$\approx 61001.96$). The $JWST$/NIRCam F277W band has the longest
observed wavelength of any of our filters, the smallest predicted
extinction at $z_{\mathrm{lens}} = 0.375$
($A_{\mathrm{F277W}} \approx 0.04$\,mag for $E(B-V) = 0.1$ mag), and
samples the slowly evolving rest-frame near-infrared
($\lambda_{\mathrm{rest}} \approx 9250\, \,$\AA). These three bands
together provide our best estimates of the flux ratios, $f_X/f_A$,  which can be compared with the macro-lens magnification
ratios $\mu_X / \mu_A$ \companioncitep[see][]{Mortsell2026} and a self-consistency check between
{$HST$} and $JWST$. 


The F150W and F160W flux ratio estimates agree to within $\sim 15$\%, providing
a useful cross-check between the two instruments. The F277W
estimates are systematically larger than both F150W and F160W by a
further $\sim 10$--$45$\, \%; we attribute this to a combination of
residual galaxy contamination, phase differences (the bluer rest-frame UV and optical
emission, dominating F115W/F150W/F160W, evolves faster than the
rest-NIR continuum sampled by F277W) and the potential differential extinction
inferred toward image~A. The latter effects suppress image~A
preferentially at bluer wavelengths and inflate the corresponding
flux ratios. F115W and F150W are potentially the most robust single-band
estimators of flux rations and thus magnification ratios, $\mu_X / \mu_A$, but should still be regarded as
carrying a $\sim 10$\,\% systematic uncertainty from these effects
in addition to the photometric uncertainty in
Table~\ref{tab:flux_ratios_by_image}. A definitive determination of the
flux ratios and inference of differential extinction between the supernova images will require final templates of the system, that is, a deep image after the SN has faded below detection that can be used to subtract from the photometry of the live SN.

\begin{deluxetable}{lccc}
\tablecaption{Relative astrometry of the SN\,2025wny system: positions of the lensed SN images (B--E), the point source S6, and the lens galaxies G1 and G2 relative to the brightest SN image A, averaged over all HST and JWST filters and epochs. The quoted uncertainties are the scatter of the individual filter/epoch measurements around the mean, adopted as the systematic uncertainty (formal statistical errors are $\lesssim$1.5\,mas).\label{tab:astrometry}}
\tablehead{
\colhead{Object} & \colhead{$\Delta\alpha\cos\delta$} & \colhead{$\Delta\delta$} & \colhead{Separation} \\
\colhead{} & \colhead{(arcsec)} & \colhead{(arcsec)} & \colhead{(arcsec)}
}
\startdata
A\tablenotemark{a} & $\equiv 0$ & $\equiv 0$ & \nodata \\
B & $0.6003 \pm 0.0016$ & $-2.1246 \pm 0.0015$ & $2.208$ \\
C & $-1.4072 \pm 0.0012$ & $-2.9136 \pm 0.0016$ & $3.236$ \\
D\tablenotemark{b} & $-3.9177 \pm 0.0033$ & $-0.2601 \pm 0.0011$ & $3.926$ \\
E & $-1.8708 \pm 0.0016$ & $1.2237 \pm 0.0026$ & $2.235$ \\
S6 & $-1.9042 \pm 0.0066$ & $-1.3212 \pm 0.0065$ & $2.318$ \\
G1 & $-1.2126 \pm 0.0024$ & $-1.3182 \pm 0.0048$ & $1.791$ \\
G2 & $-2.1794 \pm 0.0020$ & $0.8082 \pm 0.0017$ & $2.324$
\enddata
\tablenotetext{a}{Adopted position of the reference image A: R.A.\ $= 07\fh16\fm34\fs506$, Decl.\ $= +38\fdg21\farcm08\farcs12$ (J2000; $109.143776\degr$, $+38.352257\degr$), with an absolute uncertainty of $\pm0.026$ (R.A.) and $\pm0.011$ (Decl.) arcsec, dominated by the visit-level WCS calibration. The relative offsets in this table are independent of this absolute calibration.}
\tablenotetext{b}{The R.A.\ offset differs by $6.8$\,mas between HST/WFC3-UVIS and JWST/NIRCam.}
\tablecomments{SN images and S6: inverse-variance-weighted means of 11 filter/epoch measurements ($HST$: F475W, F625W, F814W, F160W; $JWST$: F115W, F150W, F277W). G1 (primary lens galaxy) and G2 (secondary galaxy): S\'ersic-fit centroids transformed to offsets within each frame, averaged after excluding the discrepant F160W 2025-11-30 fit.}
\end{deluxetable}

\subsection{Spectroscopic properties}

\begin{figure*}
    \centering
    \includegraphics[width=\textwidth]{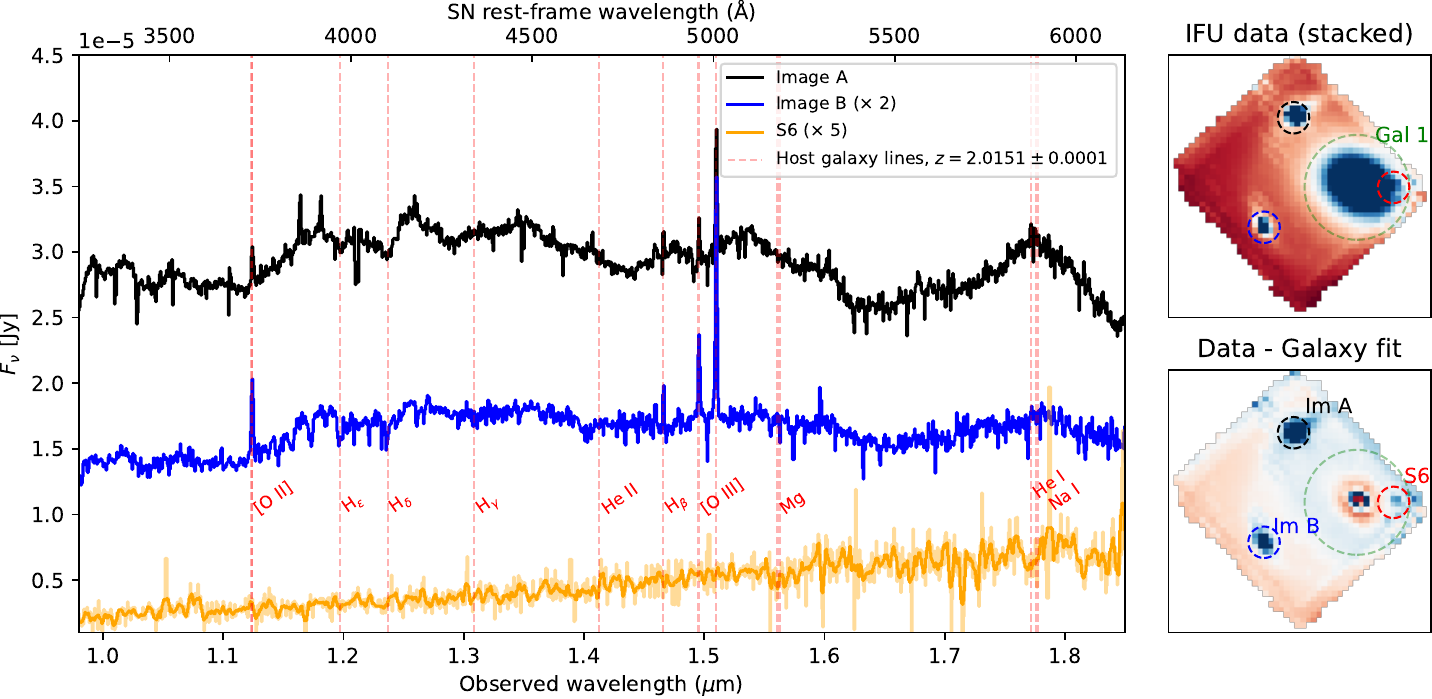}
    \caption{$JWST$/NIRSpec IFU spectroscopy of SN~2025wny. \textit{Left:} 
Extracted rest-frame spectra of lensed images A (black) and B (blue, 
scaled by a factor of 2 for clarity) and the point source S6 (orange, 
scaled by a factor of 5), obtained with the G140M/F100LP 
disperser--filter combination. The observed wavelength range 
$1.0$--$1.89\,\mu$m corresponds to rest-frame UV/optical wavelengths 
$3300$--$6300\,$\AA\ at $z=2.0151$. Prominent rest-frame galaxy lines are marked with vertical dashed lines. No statistically significant consistent set of hydrogen broad supernova features are identified, compatible with the hydrogen-poor (Type~I) 
superluminous supernova classification. \textit{Right:} Spatial 
view of the IFU data. The upper panel shows the stacked IFU 
datacube collapsed along the wavelength axis, with the position 
of lens galaxy G1 indicated. The lower panel shows the same 
collapsed image after subtraction of a S\'{e}rsic profile fit to 
G1 at each wavelength slice, revealing the resolved lensed images 
A and B and the point source S6. }
    \label{fig:ifu_spectra}
\end{figure*}

Figure~\ref{fig:ifu_spectra} shows the NIRSpec IFU spectra of SN~2025wny 
obtained with the G140M/F100LP disperser--filter combination. 
The left panel presents the extracted rest-frame spectra of lensed SN images A and B 
and the unclassified point source S6 over the observed wavelength range 
1.0--1.89\,$\mu$m, corresponding to rest-frame UV/optical wavelengths 3300--6300\,\AA\ 
at $z = 2.0155$. 

The spectrum of image~A exhibits a blue continuum with broad absorption 
features characteristic of superluminous supernovae. Crucially, no hydrogen or helium 
features are detected across the full rest-frame wavelength range, fully consistent 
with the SLSN-I classification reported in  
\citet{2025ApJ...995L..17J}. The spectrum of image~B, scaled by a factor of 3 for 
display, is qualitatively consistent with image~A, as expected given their similar 
arrival time \companioncitep{Johansson2026}. The point source S6 (scaled by a factor of 10) 
shows a distinct and significantly redder spectral shape compared to the SN images, 
suggesting that it is unrelated to both the supernova and its host galaxy, see also discussion in \companioncite{Mortsell2026}.

Superimposed on the spectra of both Images A and B are narrow emission lines from the underlying host galaxy (marked by red dashed vertical lines in \autoref{fig:ifu_spectra}). Fitting the $H_\beta$, \ion{O}{3} $\lambda4959$ and $\lambda5007$ lines yields a host-galaxy redshift of $z_{\rm host}=2.0151\pm0.0001$. This is offset by $\sim400$ km s$^{-1}$ in the rest-frame from the previously reported redshift of $z=2.011$, inferred from narrow absorption features in ground-based spectra. See \companioncite{Qin2026} for further details.

The spectral shape of S6 differs markedly from the SN images, supporting the conclusion that it is unlikely to be another image of SN~2025wny.
It is consistent with being a background source at higher redshift, 
although a faint Milky Way foreground star cannot be excluded. A definitive 
classification will require additional observations, such as template imaging after 
the supernova has faded.

The right panels of Figure~\ref{fig:ifu_spectra} illustrate the spatial decomposition of 
the IFU datacube. The upper panel shows the datacube collapsed along the wavelength 
axis, with lens galaxy G1 clearly visible. After subtracting a S\'{e}rsic profile 
fit to G1 at each wavelength slice, the lower panel shows the two brightest lensed 
SN images A and B and the point source S6 as distinct resolved sources. This 
galaxy-subtraction procedure is essential for obtaining uncontaminated spectra of 
the individual SN images, particularly for the images which lie in close angular 
proximity to G1 and G2.

The signal-to-noise ratio achieved for the spectrum of image~A is comparable to that routinely  obtained for SLSNe at $z\lesssim 0.2$, underscoring the power of the 
gravitational lensing magnification combined with $JWST$ sensitivity.

\subsection{Properties of SN host and deflecting galaxies}
\begin{figure}
    \centering
    \includegraphics[width=\columnwidth]{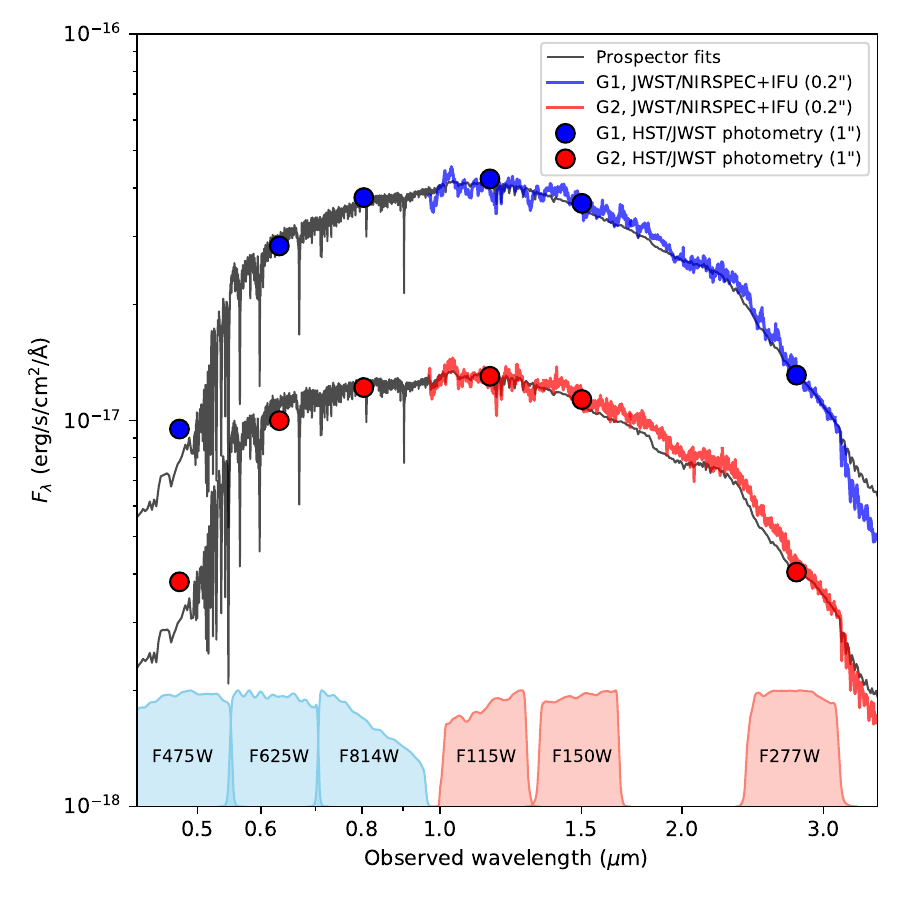}  
    \caption{Spectral energy distributions of the two foreground lensing galaxies G1 and G2. For G1, we show the \textit{JWST}/NIRSpec IFU spectrum extracted within a $0\farcs2$ aperture radius (blue solid line) and {$HST$}+{$JWST$} broadband photometry measured within a $1\farcs0$ aperture radius (blue filled circles). For G2, we show the \textit{JWST}/NIRSpec IFU spectrum extracted within a $0\farcs2$ aperture radius (red solid line) 
    and {$HST$}+{$JWST$} broadband photometry within a $1\farcs0$ aperture (red filled circles).}
\label{fig:lens_SEDs}
    \label{fig:galaxy_sed}
\end{figure}

In Figure \ref{fig:galaxy_sed}, we fit the observed photometry (Table \ref{tab:deflector_photometry}) and spectroscopy with the software package \textsc{Prospector} version 1.4 \citep{Johnson2021a}\footnote{\textsc{Prospector} uses the \textsc{Flexible Stellar Population Synthesis} (\textsc{FSPS}) code \citep{Conroy2009a} to generate the underlying physical model and \textsc{python-fsps} \citep{ForemanMackey2014a} to interface with \textsc{FSPS} in \textsc{python}. The \textsc{FSPS} code also accounts for the contribution of the diffuse gas using the \textsc{Cloudy} models of \citet{Byler2017a}. We use the dynamic nested sampling package \textsc{dynesty} \citep{Speagle2020a} to sample the posterior probability.}, assuming a Chabrier IMF \citep{Chabrier2003a} and
a star-formation history of functional form $t \times \exp\left(-t/t_{1/e}\right)$, where $t$ is the age of the star-formation episode and $t_{1/e}$ is the $e$-folding timescale, 
and the \citet{Calzetti2000} attenuation model. The priors of the model parameters are set identically to those used by \citet{Schulze2021a}.
We measure stellar masses of $\log_{10}(M_\star/M_\odot) = 11.31^{+0.13}_{-0.17}$ and 
$10.67^{+0.17}_{-0.18}$ for G1 and G2, respectively. The stellar mass of G1 is consistent with typical lenses in the SLACS sample \citep{Auger_2009}. The stellar mass of G2 is instead closer to those inferred for the lens galaxies of iPTF16geu and SN\,Zwicky  \citep{goobar_2023}.

The dominance of G1 in stellar mass is consistent with its role as the primary deflector \companioncitep[see][]{Mortsell2026}. 

Table \ref{tab:arc_phot} summarizes the host-galaxy arc space-based photometry, which is further discussed in \companioncitep{Qin2026}.

\begin{deluxetable*}{lllcccccc}
\tablecaption{HST and JWST PSF photometry of the lensed images of SN\,2025wny (A--E) and the point source S6. Magnitudes are in the AB system; uncertainties include the formal fit error, the PSF-shape uncertainty, and the fake-source injection scatter, added in quadrature. Upper limits correspond to S/N $<$ 3.
\label{tab:photometry}}
\tablehead{
\colhead{Date} & \colhead{Instrument} & \colhead{Filter} & \colhead{A} & \colhead{B} & \colhead{C} & \colhead{D} & \colhead{E} & \colhead{S6} \\
\colhead{(UT)} & \colhead{} & \colhead{} & \colhead{(mag)} & \colhead{(mag)} & \colhead{(mag)} & \colhead{(mag)} & \colhead{(mag)} & \colhead{(mag)}
}
\startdata
2025-10-20 & HST/WFC3-UVIS & F475W & $20.64 \pm 0.02$ & $22.12 \pm 0.02$ & $22.25 \pm 0.02$ & $22.71 \pm 0.02$ & $23.85 \pm 0.03$ & $>27.1$ \\
2025-10-20 & HST/WFC3-UVIS & F625W & $20.00 \pm 0.01$ & $21.64 \pm 0.01$ & $21.76 \pm 0.02$ & $22.68 \pm 0.02$ & $23.45 \pm 0.05$ & $>26.5$ \\
2025-10-20 & HST/WFC3-UVIS & F814W & $19.79 \pm 0.01$ & $21.46 \pm 0.01$ & $21.59 \pm 0.02$ & $22.52 \pm 0.03$ & $23.22 \pm 0.05$ & $25.36 \pm 0.33$ \\
2025-10-20 & HST/WFC3-IR & F160W & $20.30 \pm 0.02$ & $21.58 \pm 0.03$ & $21.85 \pm 0.04$ & $22.44 \pm 0.04$ & $23.71 \pm 0.16$ & $>24.4$ \\
\hline
2025-11-22 & JWST/NIRCam & F115W & $20.71 \pm 0.01$ & $22.25 \pm 0.01$ & $22.27 \pm 0.01$ & $23.38 \pm 0.02$ & $23.83 \pm 0.04$ & $24.91 \pm 0.09$ \\
2025-11-22 & JWST/NIRCam & F150W & $20.65 \pm 0.01$ & $22.16 \pm 0.01$ & $22.32 \pm 0.01$ & $23.24 \pm 0.02$ & $23.99 \pm 0.04$ & $24.58 \pm 0.08$ \\
2025-11-22 & JWST/NIRCam & F277W & $20.87 \pm 0.01$ & $22.26 \pm 0.01$ & $22.63 \pm 0.02$ & $23.02 \pm 0.02$ & $24.05 \pm 0.09$ & $24.12 \pm 0.07$ \\
\hline
2025-11-30 & HST/WFC3-UVIS & F475W & $20.93 \pm 0.02$ & $22.18 \pm 0.02$ & $22.39 \pm 0.02$ & $22.76 \pm 0.02$ & $24.06 \pm 0.03$ & $>27.4$ \\
2025-11-30 & HST/WFC3-UVIS & F625W & $20.58 \pm 0.01$ & $21.93 \pm 0.02$ & $22.14 \pm 0.02$ & $22.75 \pm 0.02$ & $23.82 \pm 0.05$ & $>26.1$ \\
2025-11-30 & HST/WFC3-UVIS & F814W & $20.37 \pm 0.01$ & $21.84 \pm 0.02$ & $22.02 \pm 0.02$ & $22.80 \pm 0.03$ & $23.59 \pm 0.06$ & $>25.5$ \\
2025-11-30 & HST/WFC3-IR & F160W & $20.46 \pm 0.02$ & $21.74 \pm 0.04$ & $21.92 \pm 0.05$ & $22.67 \pm 0.07$ & $>23.9$ & $>24.2$
\enddata
\end{deluxetable*}

\begin{deluxetable*}{lccccccc}
\tabletypesize{\small}
\tablecaption{Flux Ratios of Lensed SN\,2025wny Images Relative to Image~A\label{tab:flux_ratios_by_image}}
\tablecolumns{8}
\tablehead{
  \colhead{Image} &
  \colhead{$f_X/f_A$ (F475W)} &
  \colhead{$f_X/f_A$ (F625W)} &
  \colhead{$f_X/f_A$ (F814W)} &
  \colhead{$f_X/f_A$ (F115W)} &
  \colhead{$f_X/f_A$ (F150W)} &
  \colhead{$f_X/f_A$ (F160W)} &
  \colhead{$f_X/f_A$ (F277W)}
}
\startdata
B & $0.2707 \pm 0.0044$ & $0.2406 \pm 0.0033$ & $0.2307 \pm 0.0030$ & $0.2423 \pm 0.0026$ & $0.2491 \pm 0.0029$ & $0.3075 \pm 0.0090$ & $0.2773 \pm 0.0037$ \\
C & $0.2373 \pm 0.0040$ & $0.2098 \pm 0.0030$ & $0.2009 \pm 0.0029$ & $0.2393 \pm 0.0027$ & $0.2160 \pm 0.0027$ & $0.2483 \pm 0.0083$ & $0.1971 \pm 0.0033$ \\
D & $0.1588 \pm 0.0026$ & $0.0982 \pm 0.0017$ & $0.0880 \pm 0.0018$ & $0.0857 \pm 0.0019$ & $0.0925 \pm 0.0019$ & $0.1363 \pm 0.0051$ & $0.1376 \pm 0.0028$ \\
E & $0.0536 \pm 0.0013$ & $0.0450 \pm 0.0015$ & $0.0453 \pm 0.0016$ & $0.0567 \pm 0.0019$ & $0.0464 \pm 0.0018$ & $0.0427 \pm 0.0060$ & $0.0534 \pm 0.0046$ \\
\enddata
\tablecomments{Flux ratios $f_X/f_A$ are inverse-variance-weighted averages over all available epochs.
HST optical filters (F475W, F625W, F814W) and F160W use two epochs (2025 Oct~20 and Nov~30);
JWST filters (F115W, F150W, F277W) use a single epoch (2025 Nov~22).
Uncertainties are the weighted combination of per-epoch photometric errors.
Image D is observed at a different supernova phase than images A, B, C, E and is included
here for completeness but excluded from the differential extinction analysis.}
\end{deluxetable*}

\begin{deluxetable}{l l c c}
\tablecaption{AB magnitudes of the two deflecting galaxies, G1 and G2, in the
SN2025wny field. Uncertainty of approximately $0.03$\,mag applies to
all values, except for F160W, with about 3$\times$ larger uncertainty.}
\label{tab:deflector_photometry}
\tablehead{
  \colhead{Instrument} & \colhead{Filter} & \colhead{G1} & \colhead{G2}
}
\startdata
HST/WFC3-UVIS & F475W & 21.76 & 22.75 \\
HST/WFC3-UVIS & F625W & 19.96 & 21.09 \\
HST/WFC3-UVIS & F814W & 19.12 & 20.35 \\
HST/WFC3-IR   & F160W & 17.95 & 19.19 \\
JWST/NIRCam   & F115W & 18.21 & 19.49 \\
JWST/NIRCam   & F150W & 17.80 & 19.07 \\
JWST/NIRCam   & F277W & 17.58 & 18.85 \\
\enddata

\end{deluxetable}

\begin{deluxetable}{l l c}
\tablecaption{AB magnitudes of the lensed host-galaxy arc in the SN2025wny
field. \textit{HST} values are the inverse-variance weighted averages of the two
epochs (2025-10-20 and 2025-11-30); \textit{JWST} values are from the single
epoch (2025-11-22).\label{tab:arc_phot}}
\tablehead{
  \colhead{Instrument} & \colhead{Filter} & \colhead{$m_{\rm AB}$}
}
\startdata
HST/WFC3-UVIS & F475W & $24.156 \pm 0.030$ \\
HST/WFC3-UVIS & F625W & $24.071 \pm 0.074$ \\
HST/WFC3-UVIS & F814W & $24.078 \pm 0.102$ \\
HST/WFC3-IR   & F160W & $23.113 \pm 0.077$ \\
JWST/NIRCam   & F115W & $23.564 \pm 0.059$ \\
JWST/NIRCam   & F150W & $23.171 \pm 0.034$ \\
JWST/NIRCam   & F277W & $22.940 \pm 0.030$ \\
\enddata
\end{deluxetable}

\section{Summary and Conclusions}
\label{sec:conclusions}

We have presented space-based follow-up observations of SN~2025wny, the first 
strongly gravitationally lensed superluminous supernova discovered {\citep{2025ApJ...995L..17J}, obtained 
with the \textit{Hubble Space Telescope} and the \textit{James Webb Space 
Telescope}. The combination of {$HST$}/WFC3 optical imaging at two epochs 
and {$JWST$}/NIRCam near-infrared imaging and NIRSpec IFU spectroscopy 
provides an exceptionally detailed characterization of this remarkable system. 
Our main conclusions are as follows:

\begin{enumerate}

    \item SN~2025wny at $z=2.0155$ is resolved into five distinct lensed images (A--E) by a 
    pair of massive foreground galaxies (G1 and G2), $z_{\rm{G1}} = 0.3755$ and $z_{\rm{G2}} = 0.3766$. The angular 
    separations of the images range from $2\farcs2$ to $3\farcs9$ relative to 
    the brightest image A. This is the highest-redshift multiply imaged 
    supernova identified in a galaxy-scale lensing configuration to date.

    \item Precise relative astrometry of the five lensed images, the unclassified point 
    source S6, and the two deflecting galaxies is obtained from 
    inverse-variance-weighted averages over eleven filter and epoch combinations 
    from \textit{HST} and \textit{JWST}. The resulting image positions are 
    determined to better than $3\,$mas in most cases, providing the astrometric 
    foundation for the lens modeling presented in \companioncite{Mortsell2026}.

    \item Multi-band PSF photometry in seven filters spanning the observed 
    optical to near-infrared ($0.47$--$2.77\,\mu$m, corresponding to rest-frame 
    $1550$--$9200\,$\AA\ at $z=2.0155$) reveals that image A is the brightest and  
    most magnified image. Deviations from achromatic flux 
    ratios among the images are consistent with combination of differential dust 
    extinction in the lens galaxy along the sightline to image A, chromatic microlensing, residual galaxy contamination and intrinsic color evolution driven by  image-to-image time delays.

    \item The {$JWST$}/NIRSpec IFU spectrum of SN~2025wny exhibits a blue 
    continuum and broad absorption features with no detectable hydrogen supernova lines over the rest-frame wavelength range $3300$--$6300\,$\AA, 
    fully consistent with a hydrogen-poor (Type~I) superluminous supernova 
    classification at $z=2.0155$. The quality of the data is comparable to that 
    routinely achieved in the local universe, demonstrating the power 
    of gravitational lensing magnification combined with {$JWST$} 
     sensitivity. The spectrum of a sixth detected unresolved source (S6) is incompatible with being another image of SN~2025wny.

    \item SED fitting of the two deflecting galaxies with \textsc{Prospector} 
    yields stellar masses of $\log_{10}(M_\star/M_\odot) = 11.31^{+0.13}_{-0.17}$ and 
    $10.67^{+0.17}_{-0.18}$ for G1 and G2, respectively. Both are early-type galaxies, and are located close to the two loci found for galaxy-scale lenses discovered through static sources \citep{Auger_2009} and transient lens searches \citep{goobar_2023}. The dominance of G1 in stellar mass is consistent with 
    its role as the primary deflector.

    \item The observations presented here provide a foundation for a 
    suite of companion analyses, including lens modeling and time-delay 
    cosmography \companioncitep{Mortsell2026}, spectroscopic time delays \companioncitep{Johansson2026}, photometric time delays \companioncitep{Townsend2026}, and detailed studies 
    of the superluminous supernova physics \companioncitep{Li2026} and host galaxy 
    properties \companioncitep{Qin2026}. Together, these analyses establish SN~2025wny 
    as the first galaxy-scale lensed supernova suitable for precision 
    time-delay cosmography, opening a new observational window on both 
    the Hubble constant and the physics of stellar explosions near the 
    epoch of peak cosmic star formation.

\end{enumerate}

\section*{Acknowledgements}
Based on observations made with the NASA/ESA \textit{Hubble Space Telescope} and the \textit{James Webb Space Telescope}. Support for this work was provided by NASA through grants associated with HST and JWST programs GO 5564, GO 17611 and DDT 12510.
A.G.\ acknowledges financial support from the research project grant “Understanding the Dynamic Universe” funded by the Knut and Alice Wallenberg under Dnr KAW 2018.0067, and from {\em Vetenskapsr\aa det}, the Swedish Research Council through grants projects Dnr 2020-03444 and 2025-03692 the G.R.E.A.T research environment, Dnr 2016-06012, as well as the EDUCATE excellence center
funded by the Swedish Research Council through grant Dnr 2022-06627 and the Swedish National Space Agency, Dnr 2023-00226. 
E.M. acknowledges support from the Swedish Research Council under Dnr VR 2024-03927. 
 S.D. and A.T., acknowledge support from  UK Research and Innovation (UKRI) under the UK government’s Horizon Europe funding Guarantee EP/Z000475/1.

S.T. has been supported by funding from the European Research Council (ERC) under the European Union's Horizon 2020 research and innovation programmes (grant agreement no. 101018897 CosmicExplorer).

PEN used resources of the National Energy Research Scientific Computing Center (NERSC), a Department of Energy User Facility using NERSC award DDR-ERCAP-37654.

GPS acknowledges support from the Science and Technology Facilities Council (grant number ST/X001296/1)

N.A. is supported by the COFUND action of Horizon Europe’s Marie Sklodowska-Curie Actions research programme, Grant Agreement 101081355 (SMASH).


\appendix

\renewcommand{\thetable}{A\arabic{table}}
\setcounter{table}{0}
\section) {$HST$ and $JWST$ setups}
\begin{deluxetable*}{lll}
\tablecaption{{$JWST$}/GO~5564 observations of SN 2025wny.
\label{tab:jwst_obs}}
\tablehead{
\colhead{Parameter} &
\colhead{NIRCam Imaging} &
\colhead{NIRSpec IFU}
}
\startdata
Execution date (UT)      & 2025 Nov 22, 21:53:26                & 2025 Nov 19, 00:36:04 \\
V3 position angle        & $262\fdg06$                          & $264\fdg08$ \\
Template                 & NIRCam Imaging                       & NIRSpec IFU Spectroscopy \\
Module / aperture        & ALL, FULL subarray                   & IFU, $3\farcs0\times3\farcs0$ \\
Filter / disperser       & \textit{F115W}+\textit{F277W};       & \textit{G140M} / \textit{F100LP} \\
                         & \textit{F150W}+\textit{F277W}        & ($R\simeq1000$, $0.97$--$1.89~\micron$) \\
Readout pattern          & \texttt{BRIGHT1}                     & \texttt{NRSIRS2RAPID} \\
$N_{\mathrm{groups}}$ / $N_{\mathrm{int}}$ & 3 / 2              & 85 / 1 \\
Primary dither           & \texttt{INTRAMODULEBOX} ($\times 2$) & \texttt{4-POINT-DITHER} \\
Subpixel dither          & \texttt{SMALL-GRID-DITHER} ($\times 2$) & \nodata \\
Science duration (s)     & 946                                  & 5019 \\
\enddata
\end{deluxetable*}

\begin{deluxetable*}{lll}
\tablecaption{{HST}/GO~17611 WFC3 imaging of SN2025wny.
\label{tab:hst-obs}}
\tablehead{
\colhead{Parameter} &
\colhead{Visit~1 (Epoch~1)} &
\colhead{Visit~2 (Epoch~2)}
}
\startdata
Execution date (UT)          & 2025 Oct~20, 18:03                     & 2025 Nov~30,13:51 \\
\multicolumn{2}{c}{{WFC3/UVIS imaging --- aperture \texttt{UVIS2-C512C-SUB}}} \\
\textit{F475W} exposure      & $42$~s, 3-pt line dither              \\ 
\textit{F625W} exposure      & $13$~s, 3-pt line dither              \\ 
\textit{F814W} exposure      & $20$~s, 3-pt line dither               \\ 
Post-flash                   & $14~e^{-}$                             \\ 
\multicolumn{2}{c}{{WFC3/IR imaging --- aperture \texttt{IRSUB512}}} \\
\textit{F160W} exposure      & \texttt{SPARS25}, $N_{\rm SAMP}=10$  \\ 
                             & ($\sim199$~s, 3-pt line dither)       \\ 
\enddata
\end{deluxetable*}

\renewcommand{\thefigure}{A\arabic{figure}}
\setcounter{figure}{0}
\section{Filter-by-filter imaging of SN~2025wny}

Figures~\ref{fig:hst_475} through \ref{fig:jwst_277} present the {HST} and {$JWST$} images of SN~2025wny in the individual filters. The photometry of the individual supernova images was obtained after modelling and subtracting the contaminating light from both the lens galaxies and the lensed host galaxy, following the procedure described in \companioncitep{Mortsell2026}.

The light distribution of the primary lens galaxy (G1) was represented by the sum of a S\'ersic profile and an exponential component (corresponding to a S\'ersic profile with fixed index $n=1$), providing sufficient flexibility to reproduce both the central light profile and the extended stellar envelope. The secondary lens galaxy (G2) was modelled with a single S\'ersic profile, and a spatially uniform background component was included to account for any residual sky background.

The host galaxy of SN~2025wny was modelled in the source plane as a single S\'ersic profile. This source model was ray-traced through the best-fitting lens model to reproduce the observed host-galaxy arc in the image plane. To account for uncertainties in the lens and source models, an ensemble of reconstructions was generated from posterior samples of the lens-model parameters. For each realization, the nonlinear parameters were fixed while the linear surface-brightness amplitudes were re-optimized using \texttt{lenstronomy} \citep{Birrer2018,Birrer2021}. The final host-galaxy model was constructed by taking the median surface brightness in each image pixel over the ensemble of realizations, yielding a robust reconstruction that was subsequently subtracted prior to measuring the supernova photometry.

%
\begin{figure*}
    \centering
\includegraphics[width=0.8\textwidth]{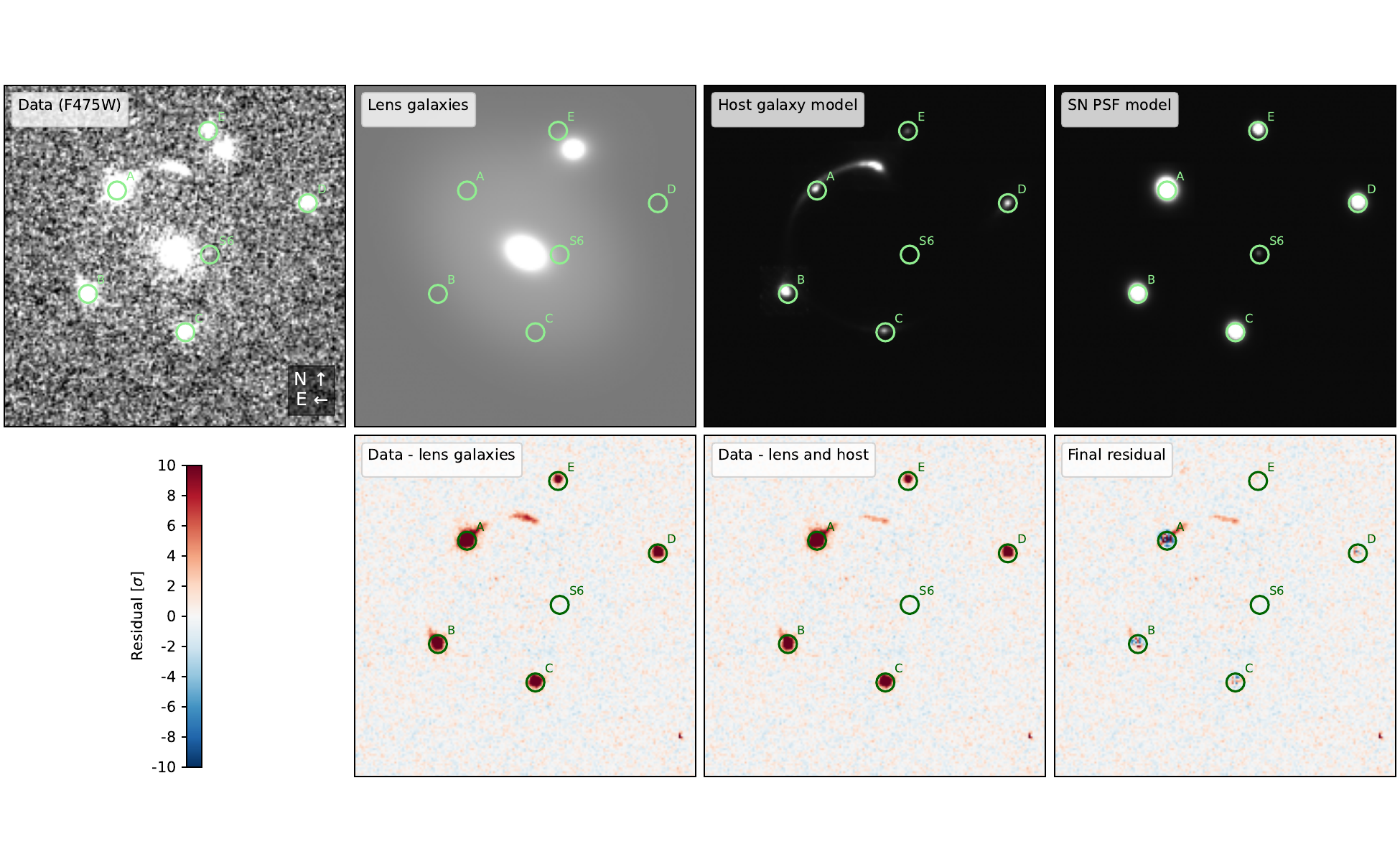}  
    \caption{Image decomposition of SN~2025wny in the \textit{HST}/WFC3 F475W band. 
Each panel is $10\farcs0 \times 10\farcs0$ with north up and east to the left. 
\textit{Top row, left to right:} The reduced science image; the best-fit 
foreground lens galaxy model (two S\'{e}rsic profiles); the best-fit host 
galaxy model; and the best-fit PSF model for the five lensed SN images 
(A--E) and point source S6. \textit{Bottom row, left to right:} The 
residual after subtracting the lens galaxy model; the residual after 
additionally subtracting the host galaxy model; and the final residual 
after subtracting all components. Residuals are shown in units of the 
local noise $\sigma$, with the color scale saturating at $\pm10\,\sigma$.}
    \label{fig:hst_475}
\end{figure*}

\begin{figure*}
    \centering
\includegraphics[width=0.8\textwidth]{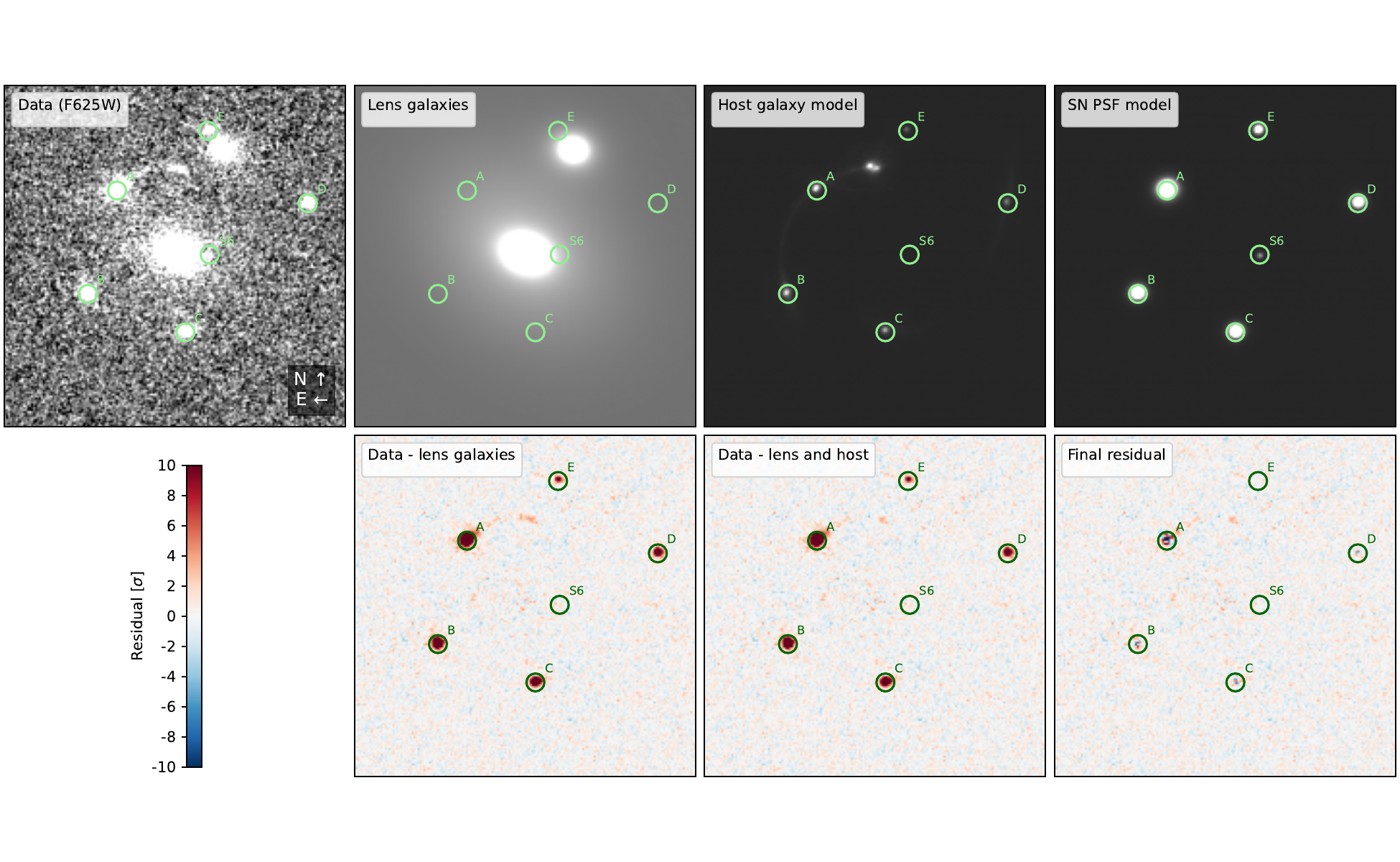}  
 \caption{Image decomposition of SN~2025wny in the \textit{HST}/WFC3 F625W band. 
Each panel is $10\farcs0 \times 10\farcs0$ with north up and east to the left. 
\textit{Top row, left to right:} The reduced science image; the best-fit 
foreground lens galaxy model (two S\'{e}rsic profiles); the best-fit host 
galaxy model; and the best-fit PSF model for the five lensed SN images 
(A--E) and point source S6. \textit{Bottom row, left to right:} The 
residual after subtracting the lens galaxy model; the residual after 
additionally subtracting the host galaxy model; and the final residual 
after subtracting all components. Residuals are shown in units of the 
local noise $\sigma$, with the color scale saturating at $\pm10\,\sigma$.}
    \label{fig:jst_625}
\end{figure*}

\begin{figure*}
    \centering
\includegraphics[width=0.9\textwidth]{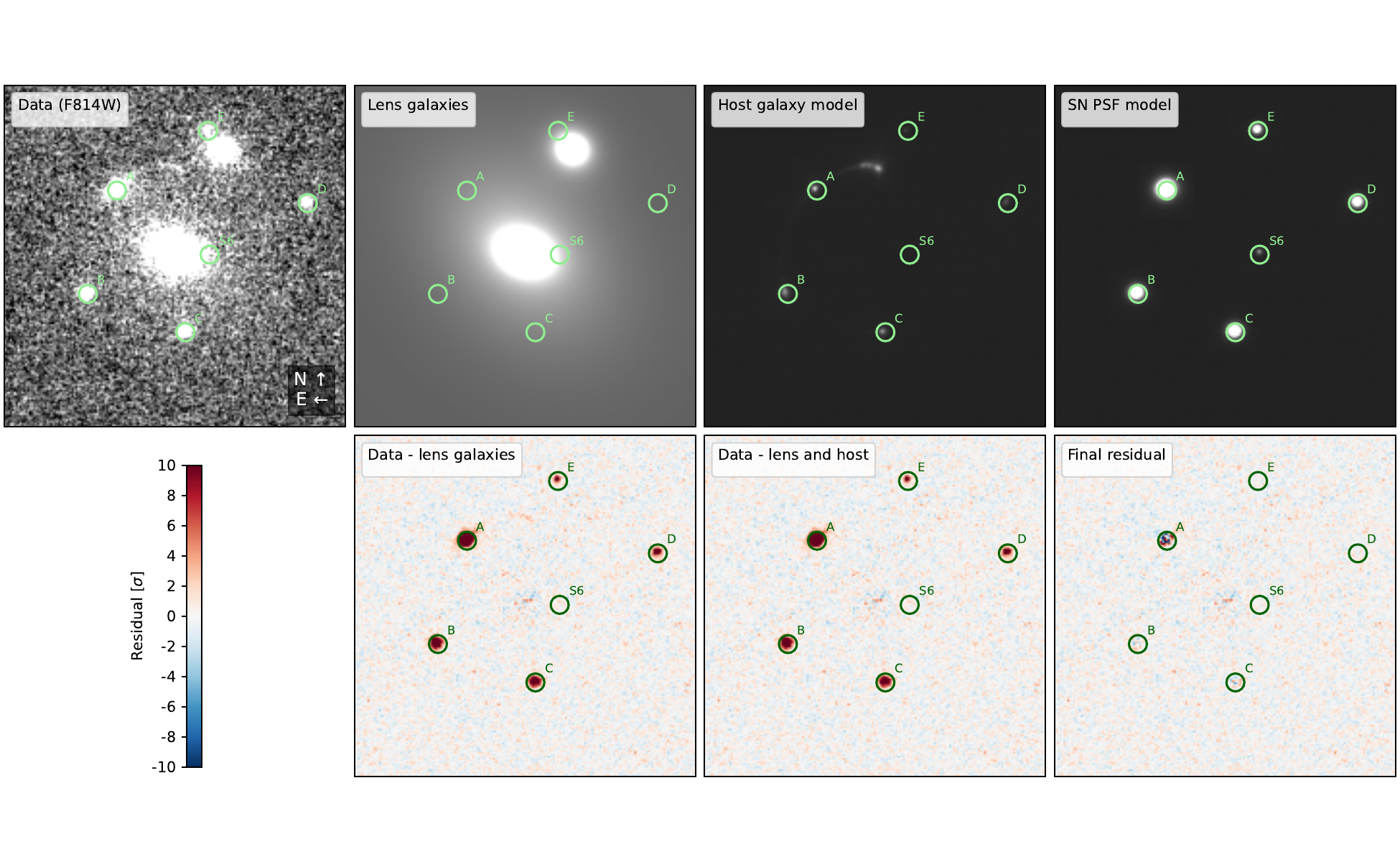}  
 \caption{Image decomposition of SN~2025wny in the \textit{HST}/WFC3 F814W band. 
Each panel is $10\farcs0 \times 10\farcs0$ with north up and east to the left. 
\textit{Top row, left to right:} The reduced science image; the best-fit 
foreground lens galaxy model (two S\'{e}rsic profiles); the best-fit host 
galaxy model; and the best-fit PSF model for the five lensed SN images 
(A--E) and point source S6. \textit{Bottom row, left to right:} The 
residual after subtracting the lens galaxy model; the residual after 
additionally subtracting the host galaxy model; and the final residual 
after subtracting all components. Residuals are shown in units of the 
local noise $\sigma$, with the color scale saturating at $\pm10\,\sigma$.}
    \label{fig:hst_814}
\end{figure*}

\begin{figure*}
    \centering
\includegraphics[width=0.9\textwidth]{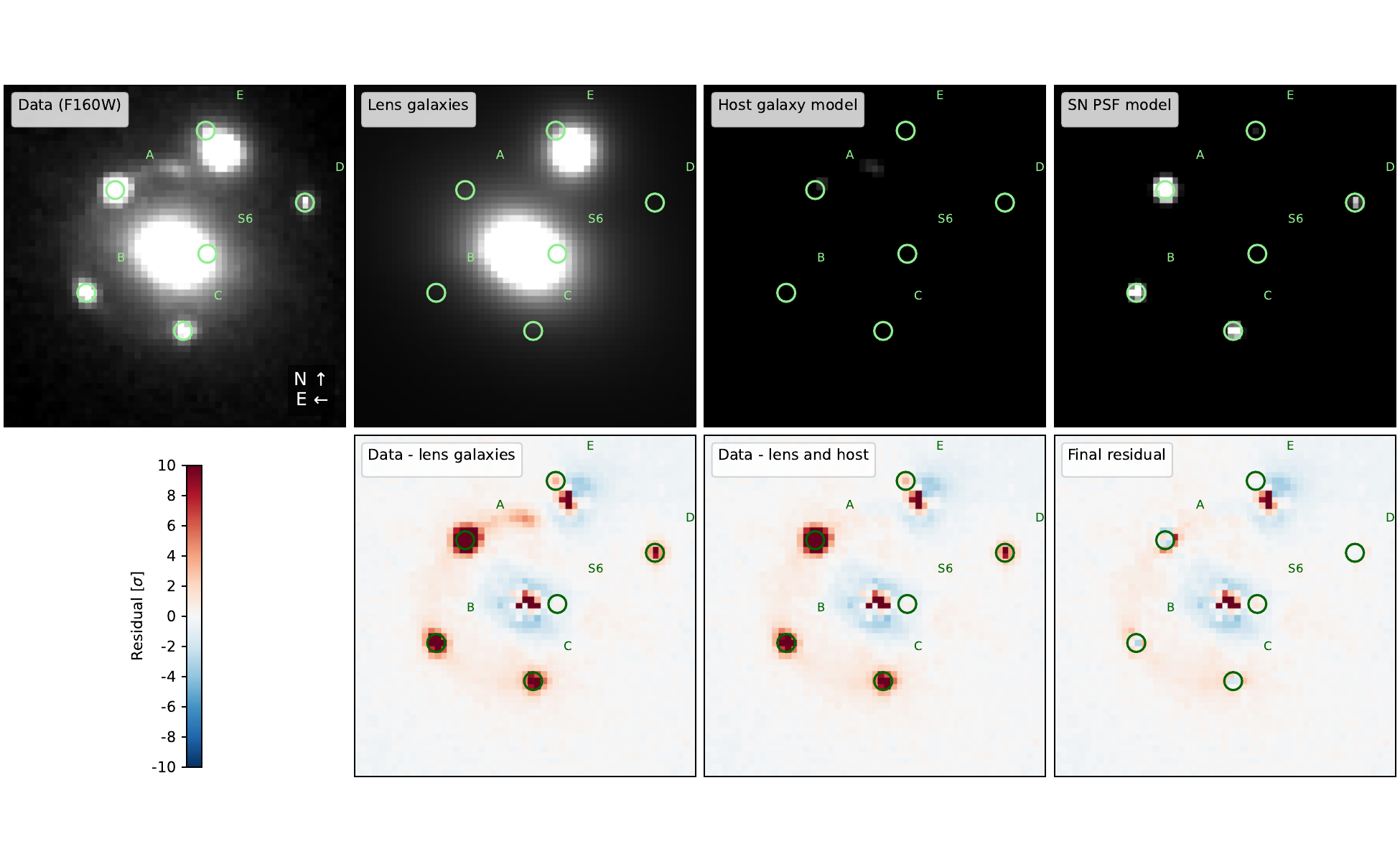}  
 \caption{Image decomposition of SN~2025wny in the \textit{HST}/WFC3 F160W band. 
Each panel is $10\farcs0 \times 10\farcs0$ with north up and east to the left. 
\textit{Top row, left to right:} The reduced science image; the best-fit 
foreground lens galaxy model (two S\'{e}rsic profiles); the best-fit host 
galaxy model; and the best-fit PSF model for the five lensed SN images 
(A--E) and point source S6. \textit{Bottom row, left to right:} The 
residual after subtracting the lens galaxy model; the residual after 
additionally subtracting the host galaxy model; and the final residual 
after subtracting all components. Residuals are shown in units of the 
local noise $\sigma$, with the color scale saturating at $\pm10\,\sigma$.}
    \label{fig:hst_160}
\end{figure*}

\begin{figure*}
    \centering
\includegraphics[width=0.9\textwidth]{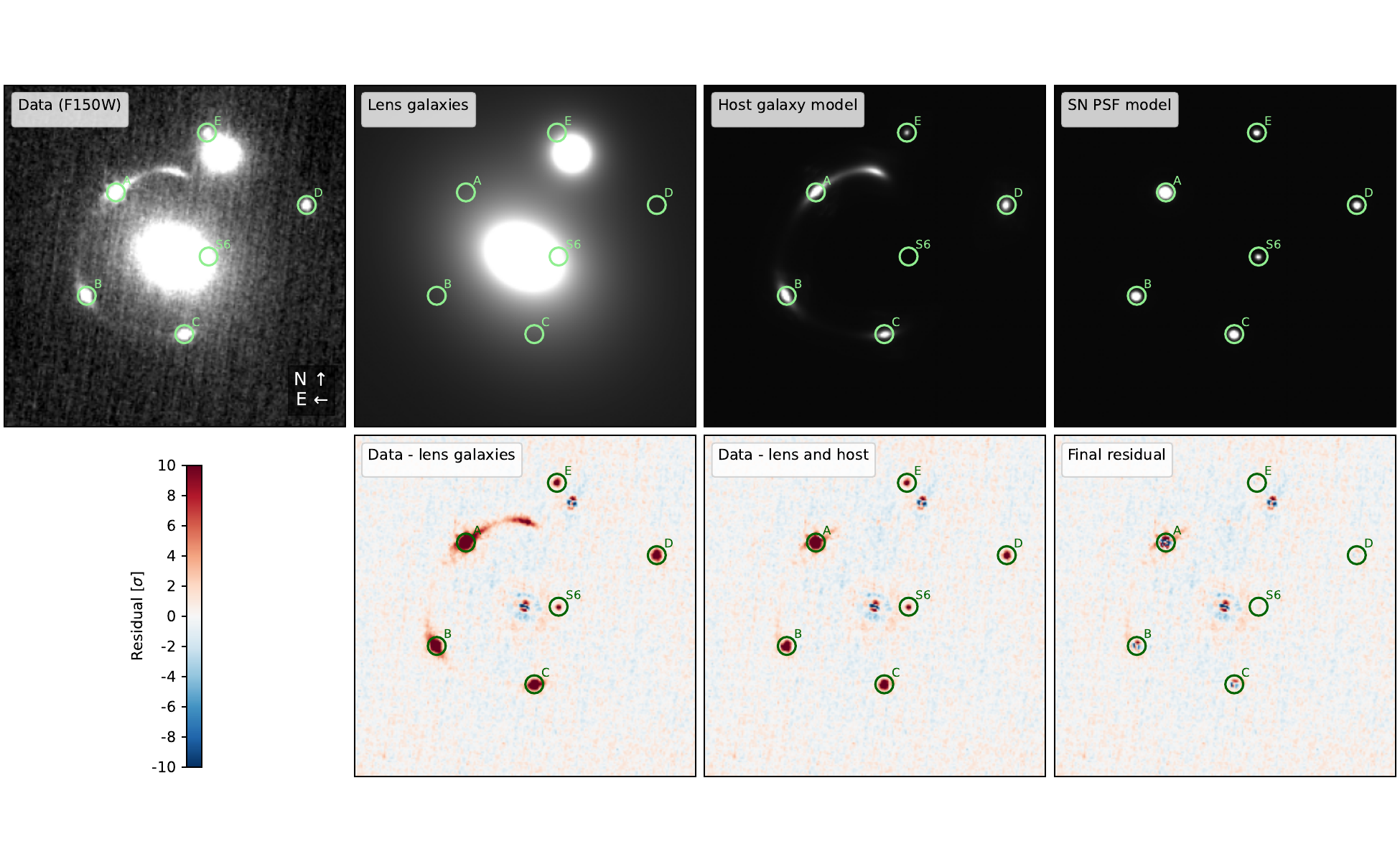}  
\caption{Image decomposition of SN~2025wny in the \textit{JWST}/NIRCam F150W band. 
Each panel is $10\farcs0 \times 10\farcs0$ with north up and east to the left. 
\textit{Top row, left to right:} The reduced science image; the best-fit 
foreground lens galaxy model (two S\'{e}rsic profiles); the best-fit host 
galaxy model; and the best-fit PSF model for the five lensed SN images 
(A--E) and point source S6. \textit{Bottom row, left to right:} The 
residual after subtracting the lens galaxy model; the residual after 
additionally subtracting the host galaxy model; and the final residual 
after subtracting all components. Residuals are shown in units of the 
local noise $\sigma$, with the color scale saturating at $\pm10\,\sigma$.}
    \label{fig:jwst_150}
\end{figure*}

\begin{figure*}
    \centering
\includegraphics[width=0.9\textwidth]{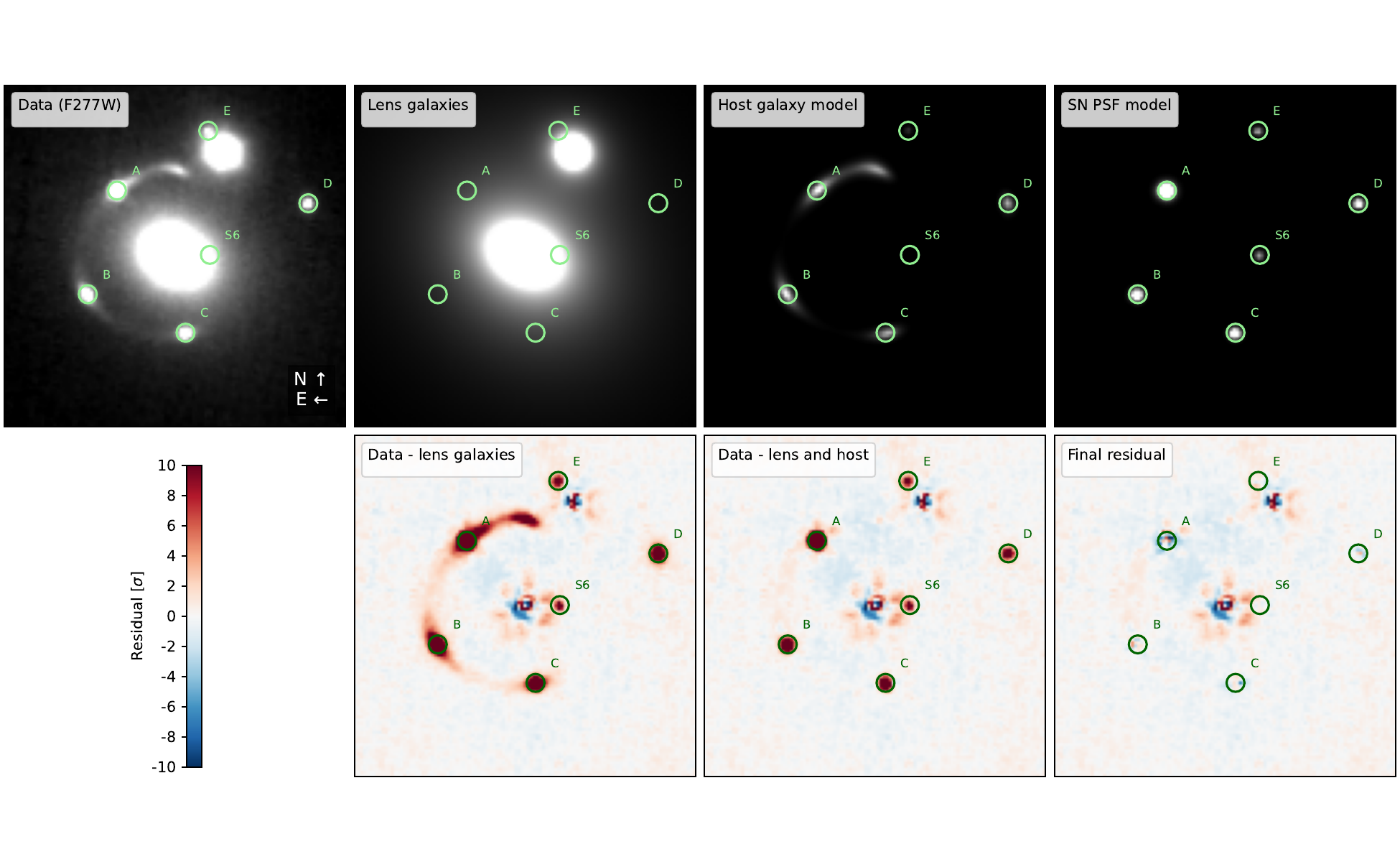}  
\caption{Image decomposition of SN~2025wny in the \textit{JWST}/NIRCam F277W band. 
Each panel is $10\farcs0 \times 10\farcs0$ with north up and east to the left. 
\textit{Top row, left to right:} The reduced science image; the best-fit 
foreground lens galaxy model (two S\'{e}rsic profiles); the best-fit host 
galaxy model; and the best-fit PSF model for the five lensed SN images 
(A--E) and point source S6. \textit{Bottom row, left to right:} The 
residual after subtracting the lens galaxy model; the residual after 
additionally subtracting the host galaxy model; and the final residual 
after subtracting all components. Residuals are shown in units of the 
local noise $\sigma$, with the color scale saturating at $\pm10\,\sigma$.}
    \label{fig:jwst_277}
\end{figure*}

\end{document}